\documentclass[twocolumn]{aa}

\usepackage{natbib}
\usepackage{amsmath}
\usepackage{graphicx}
\usepackage{xcolor}
\usepackage{multirow}
\usepackage{threeparttable}
\usepackage{booktabs}
\usepackage[breaklinks=true]{hyperref}
\usepackage[all]{hypcap}
\hypersetup{
    colorlinks,
    linkcolor={red!50!black},
    citecolor={blue!50!black},
    urlcolor={blue!80!black}
}

\begin{document}

\title{Carbon radio recombination lines from gigahertz to megahertz frequencies towards Orion A}

\author{P.~Salas\inst{1}
          \and 
          J.~B.~R. Oonk\inst{1,2}
          \and
          K.~L.~Emig\inst{1}
          \and 
          C.~Pabst\inst{1}
          \and
          M.~C.~Toribio\inst{1}
          \and
          H.~J.~A.~R\"ottgering\inst{1}
          \and
          A.~G.~G.~M.~Tielens\inst{1}}
\institute{Leiden Observatory, Leiden University, P.O. Box 9513, NL-2300 RA Leiden, The Netherlands
           \and
           Netherlands Institute for Radio Astronomy (ASTRON), Postbus 2, 7990 AA Dwingeloo, The Netherlands
           }

\abstract
{
The combined use of carbon radio recombination lines (CRRLs) and the $158$~$\mu$m-[CII] line is a powerful tool for the study of the energetics and physical conditions (e.g., temperature and density) of photodissociation regions (PDRs). 
However, there are few observational studies that exploit this synergy.
}
{Here we explore the relation between CRRLs and the $158$~$\mu$m-[CII] line in light of new observations and models.}
{
We present new and existing observations of CRRLs in the frequency range $0.15$--$230$~GHz with ALMA, VLA, the GBT,  Effelsberg 100m, and LOFAR towards Orion~A (M42). 
We complement these observations with SOFIA observations of the $158$~$\mu$m-[CII] line.
We studied two PDRs: the foreground atomic gas, known as the Veil, and the dense PDR between the HII region and the background molecular cloud.
}
{
In the Veil we are able to determine the gas temperature and electron density, which we use to measure the ionization parameter and the photoelectric heating efficiency.
In the dense PDR, we are able to identify a layered PDR structure at the surface of the molecular cloud to the south of the Trapezium cluster.
There we find that the radio lines trace the colder portion of the ionized carbon layer, the C$^{+}$/C/CO interface.
By modeling the emission of the $158$~$\mu$m-[CII] line and CRRLs as arising from a PDR we derive a thermal pressure $>5\times10^{7}$~K~cm$^{-3}$ and a radiation field $G_{0}\approx10^{5}$ close to the Trapezium.
}
{
This work provides additional observational support for the use of CRRLs and the $158$~$\mu$m-[CII] line as complementary tools to study dense and diffuse PDRs, and highlights the usefulness of CRRLs as probes of the C$^{+}$/C/CO interface.
}

\keywords{ISM: photon-dominated region (PDR) -- ISM: clouds -- radio lines : ISM -- ISM: individual objects: Orion A}

\maketitle

\section{Introduction}
\label{sec:intro}

The transfer of material from the massive reservoirs of the cold neutral medium (CNM, $T\sim80$~K) into cold molecular clouds partially regulates the star formation cycle in a galaxy \citep[e.g.,][]{Klessen2016}.
This conversion of atomic to molecular gas is intimately related to the heating and cooling processes the gas experiences.

The heating and cooling of atomic gas can be studied in photodissociation regions \citep[PDRs; e.g.,][]{Hollenbach1999}.
These are regions where the influence of far-ultraviolet (FUV) photons shape the interstellar medium (ISM) into a layered structure with hydrogen ionized or neutral close to the source of radiation and molecular farther from it.
PDRs can be found in dense and in diffuse regions; the former happen close to sites of star formation where the FUV radiation from young stars impinges on the surface of their natal molecular cloud, while the latter can be found throughout most of the CNM, powered by the interstellar radiation field (ISRF).

For the CNM and in PDRs, one of the main cooling mechanisms is through the far-infrared (FIR) fine-structure line of ionized carbon ([CII]) at $158$~$\mu$m \citep[e.g.,][]{Field1969,Dalgarno1972,Pottasch1979,Wolfire1995,Wolfire2003}.
Since carbon has a lower ionization potential than hydrogen, it is ionized throughout the diffuse ISM and in PDR surfaces, and with an energy difference between its fine structure states of $91.2$~K, it is easily excited.
However, this implies that the $158$~$\mu$m-[CII] line will also trace other phases of the ISM.
These phases include the warm ionized medium (WIM, $T\sim8000$~K), extended low density WIM \citep[ELDWIM; e.g.,][]{Heiles1994}, extended low density HII regions \citep[e.g.,][]{Goldsmith2015}, and also the surfaces of molecular clouds \citep[e.g.,][]{Visser2009,Wolfire2010}.
It has been estimated that $\sim21\%$ of the $158$~$\mu$m-[CII] line in our Galaxy is produced in the CNM and $\sim47\%$ in dense PDRs \citep{Pineda2013}.
In order to measure the cooling rate of the CNM, we must be able to isolate its contribution to the excitation of the $158$~$\mu$m-[CII] line \citep[e.g.,][]{Pabst2017}.

The separation between cold and warm gas can be done using carbon radio recombination lines \citep[CRRLs e.g.,][]{Gordon2009}.
These are lines produced when a carbon ion recombines with an electron to a large principal quantum number $n$ resulting in a Rydberg atom.
When a Rydberg atom of carbon transitions between different principal quantum numbers it will produce CRRLs, from gigahertz to megahertz frequencies depending on the $n$ levels involved.
The optical depth of the produced CRRL has a strong dependence on the gas temperature ($\tau\propto T^{-5/2}$), so CRRLs have little contamination from warm gas or HII regions.
Thus, we can use CRRLs to isolate emission from the CNM and the surfaces of molecular clouds and their contribution to the $158$~$\mu$m-[CII] line excitation.

Another property of CRRLs is that the population of carbon atoms in each energy state is determined by the gas density, temperature, and radiation field, as well as the atomic physics involved \citep[e.g.,][]{Shaver1975,Watson1980,Salgado2017a}.
Therefore, it is possible to determine the gas physical conditions by observing CRRLs at a range of frequencies and comparing this to the predicted line properties \citep[e.g.,][]{Dupree1974,Payne1989,Roshi2011,Salgado2017b,Oonk2017,Salas2018}.

Moreover, given that CRRLs have a different temperature dependency from the $158$~$\mu$m-[CII] line, we can combine both types of lines to determine the gas temperature and/or density \citep[e.g.,][]{Natta1994,Smirnov1995,Salgado2017b,Salas2017}.
This approach is particularly useful as it requires observations of a few of the faint CRRLs ($\tau\sim10^{-3}$--$10^{-4}$) instead of a large set of them to reach a similar accuracy on the derived gas properties.
However, the combined use of the $158$~$\mu$m-[CII] line and CRRLs has been performed only towards a handful of sources and using observations which do not resolve the lines in velocity and/or do not have the same angular resolution.

One of the sources that has been studied in  CRRLs and in the $158$~$\mu$m-[CII] line is the Orion star forming region.
Orion~A is a nearby giant molecular cloud that covers $\approx29$~deg$^{2}$ \citep{Maddalena1986}.
In the northern part of this cloud we can find the Orion nebula cluster \citep[ONC; e.g.,][]{Pickering1917,Sharpless1952,ODell2001}, the region of massive star formation that is closest to Earth.
The brightest stars in the ONC are the Trapezium stars \citep[M42, $(\alpha,\delta)_{\mathrm{J2000}}=(5^{\mathrm{h}}35^{\mathrm{m}}17.3^{\mathrm{s}},-5^{\circ}23^{\mathrm{m}}28^{\mathrm{s}})$, e.g.,][]{Large1981}.
The ionizing radiation from the Trapezium stars has created a HII region.
M42 lies in front of Orion~A, which makes it easier for the ionizing radiation to escape towards the observer \citep{Zuckerman1973,Balick1974a,Balick1974b}.
Behind the Trapezium stars and the HII region, Orion~A is arranged in an S-shaped structure known as the integral shaped filament \citep[ISF,][]{Bally1987}.
In front of the Trapezium stars and the HII region, there are layers of neutral gas collectively known as the Veil \citep[e.g.,][]{vanderWerf1989,Abel2004,ODell2009,vanderWerf2013,Troland2016}.
Observations of the $21$~cm-HI line at high spatial resolution ($\approx7\arcsec$) show that the gas in the Veil is composed of two spatially distinct velocity components: component A at $5.3$~km~s$^{-1}$ and component B at $1.3$~km~s$^{-1}$ \citep[][]{vanderWerf1989}.
The proximity and geometry of M42, sandwiched between a high density molecular cloud and the diffuse gas in the Veil, makes it an ideal target to study how the gas cooling rate changes between dense and diffuse gas.

The goal of this work is to re-evaluate the relation between the $158$~$\mu$m-[CII] line and CRRLs at radio frequencies in the light of new models and observations of Orion~A.
We take advantage of new large-scale maps ($\approx1$~deg$^{2}$) of the $158$~$\mu$m-[CII] line in the FIR \citep[][]{Pabst2019}.
This improves on previous comparisons that used velocity unresolved observations of the $158$~$\mu$m-[CII] line \citep{Natta1994}.
The velocity resolution of the $158$~$\mu$m-[CII] line observations in this study was of $\approx50$~km~s$^{-1}$, while in the observations of \citet[][]{Pabst2019} this is $0.2$~km~s$^{-1}$.
Additionally, we use models that describe the level population of carbon atoms including the effect of dielectronic capture \citep{Salgado2017a}.
Incorporating this effect can change the predicted CRRL intensities by a factor of two \citep[e.g.,][]{Wyrowski1997}.

This work is organized as follows.
In Section~\ref{sec:obs} we start by presenting the observations used.
We present the results obtained from these observations  in Section~\ref{sec:results}.
In Section~\ref{sec:physcond} we derive physical conditions from our results; these conditions  are also compared against results found in the literature.
We conclude with a summary of our work in Section~\ref{sec:summary}.

In this work, all velocities are given in the local standard of rest unless otherwise specified.
To convert to heliocentric velocities $18.1$~km~s$^{-1}$ should be added.
We adopt a distance of $414$~pc to Orion~A \citep[e.g.,][]{Menten2007,Zari2017}.

\section{Observations and data reduction}
\label{sec:obs}

We start by describing previously unpublished CRRL observations.
They include an L-band ($1$~GHz to $2$~GHz) map of CRRLs; pointings towards M42, which include CRRLs at frequencies between $2.8$~GHz and $275$~MHz; and a cube of CRRL absorption at $150$~MHz.
We also briefly describe CRRL observations taken from the literature, as well as observations of other tracers relevant for this work. 

\subsection{GBT observations}

\subsubsection{L-band CRRL maps}

We observed Orion~A with the National Radio Astronomy Observatory (NRAO) Robert C. Byrd Green Bank Telescope\footnote{The Green Bank Observatory is a facility of the National Science Foundation operated under cooperative agreement by Associated Universities, Inc.} (GBT) during seven nights in November $2016$ (project: AGBT16B\_225).
We mapped a $\approx0.4\degr\times1\degr$ region centered on $(\alpha,\delta)_{\mathrm{J2000}}=5^{\mathrm{h}}35^{\mathrm{m}}14.5^{\mathrm{s}},-5\degr22^{\mathrm{m}}29.3^{\mathrm{s}}$ using the on-the-fly imaging technique \citep[e.g.,][]{Mangum2007}.
The observations were performed using the L-band ($1.1$--$1.8$~GHz) receiver and the Versatile GBT Astronomical Spectrometer \citep[VEGAS;][]{Bussa2012}.
VEGAS was set up to process $27$ spectral windows $23.44$~MHz wide.
Each spectral window was split into $65536$ channels of $0.357$~kHz in width.
The spectral windows were centered on the $21$~cm-HI line, the four $18$~cm-OH lines, and the remaining on RRLs within the GBT L-band range. 
As an absolute flux calibrator we observed 3C123 \citep{Baars1977} with the \citet{Perley2013} flux scale to convert from raw counts to temperature.
We adopted the methods described in \citet{Winkel2012} to convert the raw units to temperature when possible.
As is discussed below, in some steps this was not possible due to the high continuum brightness of Orion~A.
A summary of the observational setup is presented in Table~\ref{tab:gbt1}.

\begin{table}
\begin{threeparttable}
\caption{GBT mapping observation parameters}
\begin{tabular}{lc}
\hline
\hline
Project code                                                         & AGBT16B\_225 \\
\multirow{3}{*}{Observation dates}                                   & 5, 6, 8, 9, 10, 15 and \\
                                                                     & 17 of November 2016 \\
                                                                     & and 2 of December 2016. \\
Polarizations                                                        & XX,YY \\
Spectral windows                                                     & 27 \\
\multirow{9}{*}{\shortstack[l]{Spectral window\\frequencies (MHz)}}  & 1156, 1176, 1196,\\  
                                                                     & 1217, 1240, 1259, \\
                                                                     & 1281, 1304, 1327, \\
                                                                     & 1351, 1375, 1400, \\
                                                                     & 1420.4, 1425, 1451, \\ \\
                                                                     & 1478, 1505, 1533, \\
                                                                     & 1561, 1591, 1608, \\
                                                                     & 1621, 1652, 1665.4,\\
                                                                     & 1684, 1716, 1720.53 \\
Spectral window                                                      & \multirow{2}{*}{23.44} \\
 bandwidth (MHz)                                                     & \\
Channels per spectral window                                         & 65536 \\
Integration time                                                     & \multirow{2}{*}{10.57} \\ 
per spectral dump (s)                                                &  \\
Absolute flux calibrator                                             & 3C123 \\
Total observing time                                                 & $15$ hours\\
Principal quantum numbers\tablefootmark{a}                           & $156$--$178$ \\
\hline
\end{tabular}
\tablefoot{ 
\tablefoottext{a}{For C$n\alpha$ lines.}
}
\label{tab:gbt1}
\end{threeparttable}
\end{table}

\begin{figure}
 \includegraphics[width=0.5\textwidth]{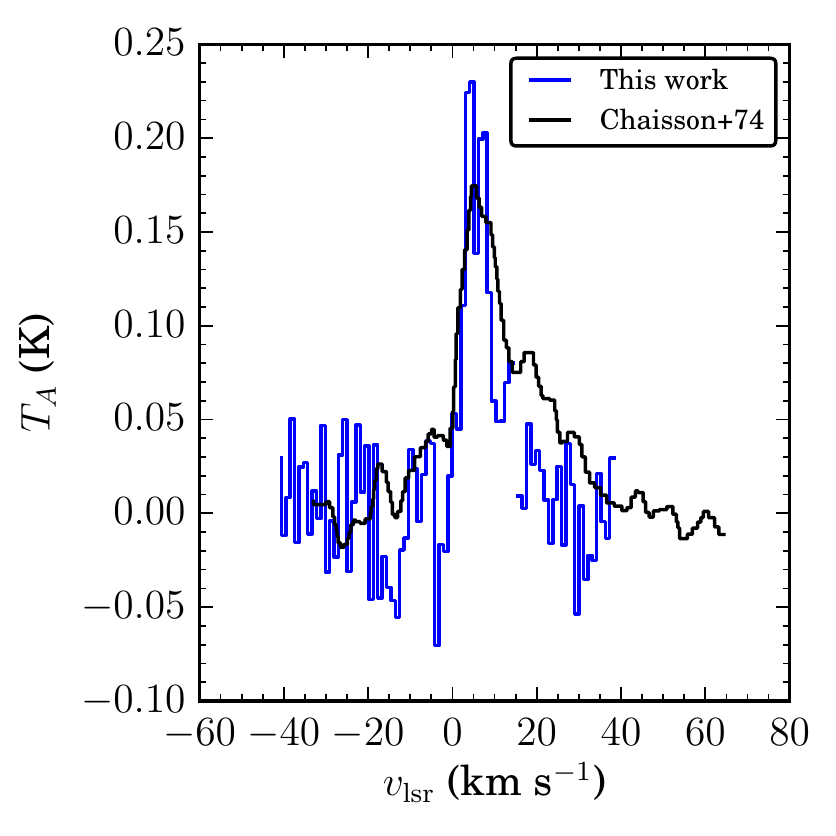}
 \caption{Comparison between the temperature-calibrated GBT C$157\alpha$ line and the C$158\alpha$ observation of \citet{Chaisson1974a}.
          The temperature scales of the two spectra agree over the brightest portion of M42, the most affected by a nonlinearity in the signal path (see text for details).
          Both spectra were extracted from an aperture of diameter $18\arcmin$ centered on $(\alpha,\delta)_{\mathrm{J2000}}=(5^{\mbox{h}}35^{\mbox{m}}17.45^{\mbox{s}},-5\degr23^{\mbox{m}}46.8^{\mbox{s}})$.
          The on-source time of the C$157\alpha$ observations is roughly seven minutes, while that of the C$158\alpha$ observations is $1200$ minutes.
          The  resolution of the C$158\alpha$ spectrum is $1.9$~km~s$^{-1}$ \citep{Chaisson1974a} and that of the C$157\alpha$ spectrum is $1$~km~s$^{-1}$.
          \label{fig:gbt_temp}}
\end{figure}

The mapped region was subdivided into smaller maps in order to keep the variations in antenna temperature within the three dB dynamical range of VEGAS.
At the beginning of each session, the pointing and focus solutions were updated on 3C161.
The pointing corrections were less than $10\%$ of the beam width.

Given the high continuum brightness of Orion~A \citep[$\sim375$~Jy or $600$~K at $1.4$~GHz, e.g.,][]{Goudis1975}, the telescope amplifiers were saturated over the brightest portions of the source.
The saturation produced a compression of the amplifier gain.
In order to correct for the  nonlinearity in the conversion from raw counts to brightness temperature in the affected portions of the map, we followed a procedure similar to that used by the GBT intermediate frequency nonlinearity project\footnote{www.gb.nrao.edu/$\sim$tminter/1A4/nonlinear/nonlinear.pdf} and briefly outlined in Appendix~\ref{app:nlg}.
To quantitatively determine the deviation from a linear gain we compared the raw counts against the $21$~cm continuum maps of \citet{vanderWerf2013}.
To scale the temperature of the \citet{vanderWerf2013} map across the $700$~MHz wide frequency range used in the GBT observations we use the results of \citet{Lockman1975}.
\citet{Lockman1975} present a compilation of continuum measurements of Orion~A.
According to these measurements, the brightness temperature of the source scales as $T_{\mathrm{M42}}\propto\nu^{-1.7}$ between $1.1$~GHz and $1.8$~GHz.
After converting the data to temperature units we compared the resulting spectra against previously published values.
An example of this comparison is shown in Figure~\ref{fig:gbt_temp}, where we compare the C$158\alpha$ spectrum against that observed by \citet{Chaisson1974a} using the $42.7$~m antenna of the NRAO.
The uncertainty on the absolute flux calibration is  $\approx20\%$, considering the nonlinear gain correction.

Before gridding all maps together, we checked that the line profiles on overlapping regions agreed.
We found no significant differences among the maps.
Then all the data was gridded together using the stand-alone \emph{GBTGRIDDER}\footnote{https://github.com/nrao/gbtgridder}.
With this program we produced a CRRL cube for each line observed.

To obtain the best spatial resolution possible from these observations, we stacked the first three CRRLs observed ($156$, $157$, and $158$) in one cube.
This produced a cube with a half power beam width (HPBW) of $\approx8\farcm1$ and an average principal quantum number of $157$.
To increase the signal-to-noise ratio of the cube, we averaged in velocity to a channel width of $\approx1$~km~s$^{-1}$ (Figure~\ref{fig:gbt_temp}).

\subsubsection{Pointings towards M42}
\label{sssec:gbtm42}

We searched the NRAO archive for observations of M42.
From the available observations we used projects AGBT02A\_028, AGBT12A\_484, and AGBT14B\_233.
They correspond to single pointings of M42 with the GBT that have a spectral resolution adequate for spectral line analysis ($\approx1$~km~s$^{-1}$ spectral resolution).
A summary of these observations is provided in Table~\ref{tab:gbt2}.

\begin{table*}
\begin{center}
\begin{threeparttable}
\caption{GBT single-pointing observation parameters}
\begin{tabular}{lcccc}
\hline
\hline
Project code & Frequency ranges & Aperture efficiency  & HPBW        & Principal quantum  \\
             & (MHz)            &                      & ($\arcmin$) & numbers\tablefootmark{a} \\
\hline
\multirow{3}{*}{AGBT02A\_028} & 275--912   & 0.7        & $41$--$14$ & $193$--$287$ \\
                              & 1100--1800 & 0.7        & $11$--$7$  & $154$--$181$ \\
                              & 1800--2800 & 0.68       & $7$--$4.4$ & $133$--$153$ \\
AGBT12A\_484 & 827--837       & 0.7                     & $12$       & 199 \\
AGBT14B\_233 & 691--761       & 0.7                     & $14$       & 204--211 \\
\hline
\end{tabular}
\tablefoot{ 
\tablefoottext{a}{For C$n\alpha$ lines.}
}
\label{tab:gbt2}
\end{threeparttable}
\end{center}
\end{table*}

\begin{figure}
 \includegraphics[width=0.5\textwidth]{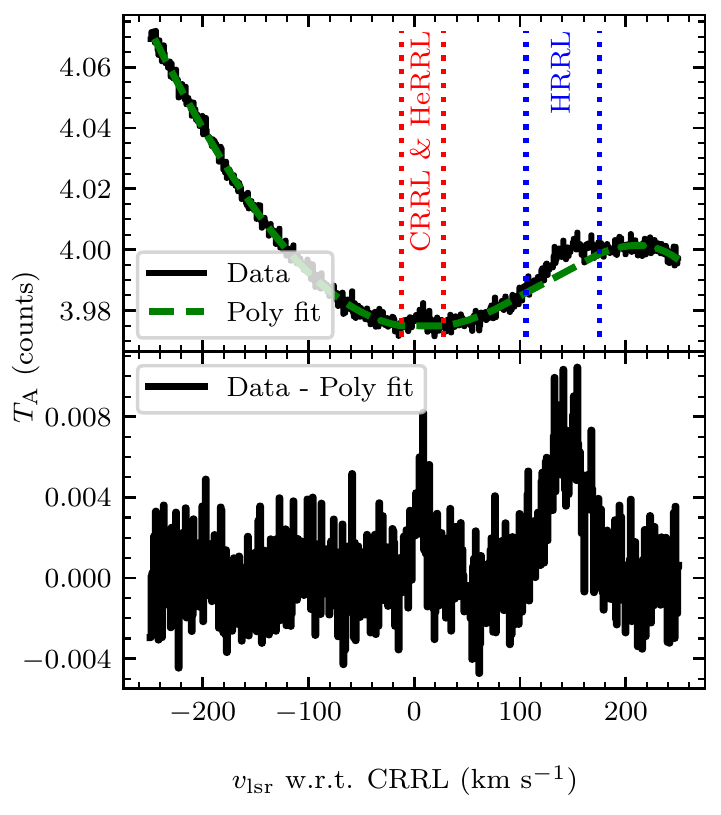}
 \caption{Example of the baseline removal process for the GBT observations.
          Upper panel: Raw data (in black) and the polynomial (green dashed line) used to remove the shape of the bandpass from the data.
          In this example a polynomial of order five was used.
          The red and blue dotted lines show the ranges where we expect the RRLs. 
          These ranges are not considered while fitting the polynomial.
          Bottom panel: Data after subtraction of the polynomial used to capture the bandpass shape.
          The velocity axis is referenced with respect to the rest frequency of the corresponding CRRL.
          This data is part of project AGBT12A\_484.
          \label{fig:gbt2red}}
\end{figure}

The data was exported to SDfits format from the NRAO archive.
The observations were calibrated to a temperature scale using the hot load on the GBT and its temperature, as listed in the SDfits header.
 To remove the continuum and any large-scale ripples in the spectra we fitted a polynomial to line-free channels.
For $97.5\%$ of the spectra an order five polynomial was used, for $2\%$ an order nine polynomial, and for the remaining an order $11$ polynomial.
If a polynomial with an order greater than $11$ was required, the data was flagged as bad and not used.
 Line-free channels are defined as those that have velocities less than $-5$~km~s$^{-1}$ and greater than $180$~km~s$^{-1}$, and those between $25$~km~s$^{-1}$ and $100$~km~s$^{-1}$, where the velocities quoted are with respect to the rest frequency of the corresponding CRRL.
An example of the polynomial fitting is shown in Figure~\ref{fig:gbt2red}.
In this example a polynomial of order five was used to remove the continuum and the  large-scale ripples.
In some cases the spectral window was flagged and marked as bad because of strong RFI.
The remaining spectra that showed no obvious artifacts were then stacked to improve the signal-to-noise ratio.

In the pointing observations present in the archive, we found no corresponding observations of a reference region.
For this reason we did not try to estimate the continuum temperature of the source from these observations.

\subsection{LOFAR observations}

We observed Orion~A with the Low Frequency Array \citep[LOFAR,][]{vanHaarlem2013} during two separate projects, two years apart.
The observations were carried out on February $2$, $2014,$ and October $27$, $2016$.
Both observations used the high band antennas (HBA) in their low  frequency range ($110$--$190$~MHz).
The number of Dutch stations available was $34$ for both observations.

Complex gain solutions were derived on 3C147 and then transferred to the target field, following a first generation calibration scheme \citep[e.g.,][]{Noordam2010}.
We adopted the \citet{Scaife2012} flux scale.
The calibrated visibilities were then imaged and cleaned.
During the inversion a Briggs weighting was used, with a robust parameter of $0$ \citep{Briggs1995}.
The cubes have a synthesized beam of $3\farcm65\times2\arcmin$ at a position angle of $166\degr$.
Given the shortest baseline present in the visibilities, $130$~m, the LOFAR observations are sensitive to emission on angular scales smaller than $53\arcmin$.

\begin{figure}
 \includegraphics[width=0.5\textwidth]{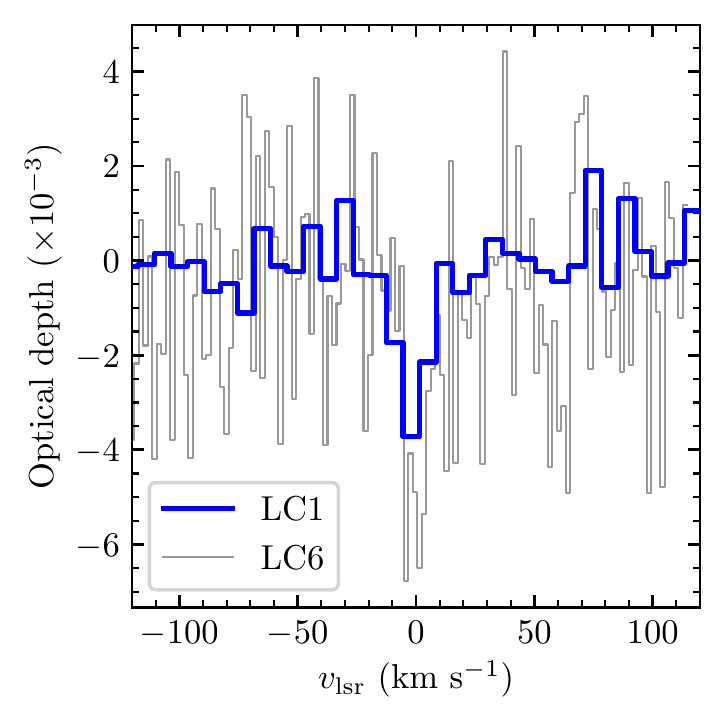}
 \caption{LOFAR spectra of C$351\alpha$ observed on two different nights.
          The blue steps show the spectra obtained from observations performed during February $2014$ and the gray steps for observations taken during October $2016$.
          The $2014$ detection, with a signal-to-noise ratio of $4.5$, is confirmed by the $2016$ observations (with a signal-to-noise ratio of $2.5$).
          The spectra are the spatial average over a $9\arcmin\times9\arcmin$ box centered on M42.
          \label{fig:c351a}}
\end{figure}

From the cubes we extracted a spectrum from a $9\arcmin\times9\arcmin$ region centered on M42.
For the $2014$ observations, $20$ spectral windows were stacked resulting in a spectrum with a spectral resolution of $7$~km~s$^{-1}$.
This resulted in a detection of the C$351\alpha$ line in absorption with a signal-to-noise ratio of $4.5$.
For the $2016$ observations, $22$ spectral windows were stacked.
In general, the data quality for the $2016$ observations was worse than in the $2014$ observations by a factor $2\mbox{--}4$.
In the $2016$ observations we found an absorption feature with a signal-to-noise ratio of $2.5$.
A comparison of the observed line profile for both observations is presented in Figure~\ref{fig:c351a}.
The line properties are consistent between the two observations.
Based on this, we are confident that the detected absorption feature, which we associate with the C$351\alpha$ line, is of astronomical origin.

\subsection{Literature data}

We  also use observations of CRRLs and other tracers of the ISM from the literature.
The CRRL observations include the C$65\alpha$ map of a $5\arcmin\times5\arcmin$ region close to M42 at a spatial resolution of $40\arcsec$, the intensity of the C$91\alpha$ line towards a region to the north of Orion-KL \citep{Wyrowski1997}, and observations of the C$30\alpha$ line using the Atacama Large Millimeter Array (ALMA) total power array plus ALMA compact array (ACA) at a spatial resolution of $28\arcsec$ \citep{Bally2017}.
Throughout this work we compare the CRRL observations with the $158$~$\mu$m-[CII] line cube observed with the Stratospheric Observatory for Infrared Astronomy \citep[SOFIA][]{Young2012} upGREAT receiver \citep[][]{Heyminck2012,Risacher2016}.
This cube has a spatial resolution of $18\arcsec$ and a velocity resolution of $0.2$~km~s$^{-1}$, and covers a region of roughly $1\degr\times1\degr$.
The observations and reduction used to produce the $158$~$\mu$m-[CII] line cube are described in detail in \citet[][]{Pabst2019}.
Additionally, we compare the CRRL cubes with observations of $^{12}$CO$(2\mbox{--}1)$ and $^{13}$CO$(2\mbox{--}1)$ \citep{Berne2014}, and with the dust properties as derived from Herschel and Planck observations \citep{Lombardi2014}.

\section{Results}
\label{sec:results}

In this section we start by describing the RRL spectra towards M42, focusing on the CRRLs.
Then we present the maps of CRRL emission that we used to study the spatial distribution of the lines and for comparison with other tracers of the ISM, particularly the $158$~$\mu$m-[CII] line.

\subsection{RRLs from M42}

\begin{figure}
 \includegraphics[width=0.5\textwidth]{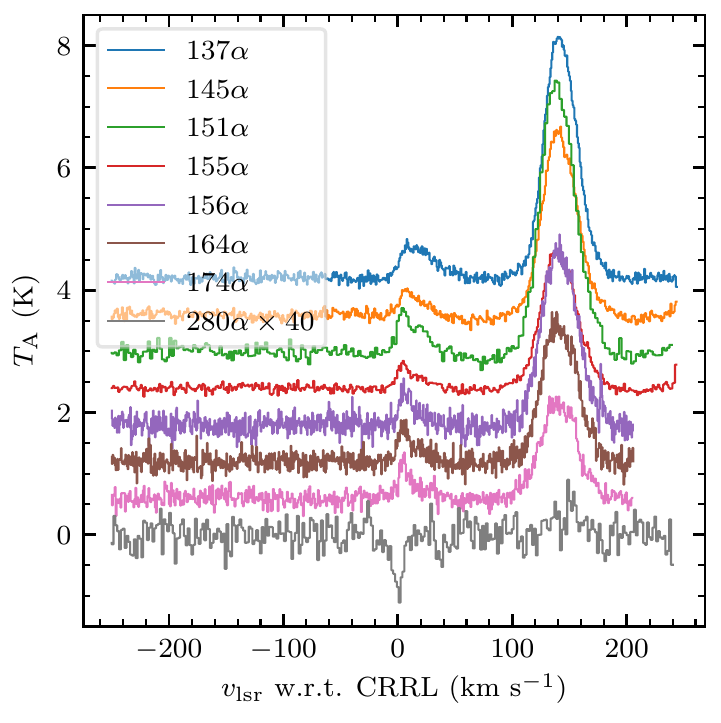}
 \caption{Hydrogen, helium, and carbon radio recombination lines observed with the GBT for $\alpha$ lines (the change in principal quantum number is $\Delta n=1$) with principal quantum numbers $137$, $145$, $151$, $155$, $156$, $164$, $174$, and $280$.
 The velocity is given with respect to the CRRL.
 To reference the velocity with respect to helium or hydrogen,   $27.4$~km~s$^{-1}$ or $149.4$~km~s$^{-1}$, respectively, is subtracted.
 The spectra are offset by a constant $0.7$~K, and the $280\alpha$ spectrum is scaled by a factor of $40$.
 This data is part of project AGBT02A\_028.
 Since all the observations are obtained using the same telescope,  their spatial resolution ranges from $4\farcm4$ to $36\arcmin$.
 \label{fig:gbt2lbandcrrls}}
\end{figure}

Some of the RRL stacks obtained from the pointed observations towards M42 are presented in Figure~\ref{fig:gbt2lbandcrrls}.
In these stacks the strongest features are hydrogen RRLs (HRRLs), followed by a blend of CRRLs and helium RRLs (HeRRLs).
The velocity difference between HeRRLs and CRRLs is  $27.4$~km~s$^{-1}$, and between HRRLs and CRRLs is $149.4$~km~s$^{-1}$.
HRRLs and HeRRLs trace the ionized gas in the HII region, for which the line FWHM due to Doppler broadening is $\approx20$~km~s$^{-1}$.
In M42 the ionized gas is blueshifted with respect to the bulk of the molecular and neutral gas \citep[e.g.,][]{Zuckerman1973,Balick1974a,Balick1974b}.
This brings the HeRRL and CRRL closer, resulting in the observed blending.
Fortuitously, we can use the fact that the HeRRLs are broader to distinguish them from the CRRLs.

\begin{figure}
 \includegraphics[width=0.5\textwidth]{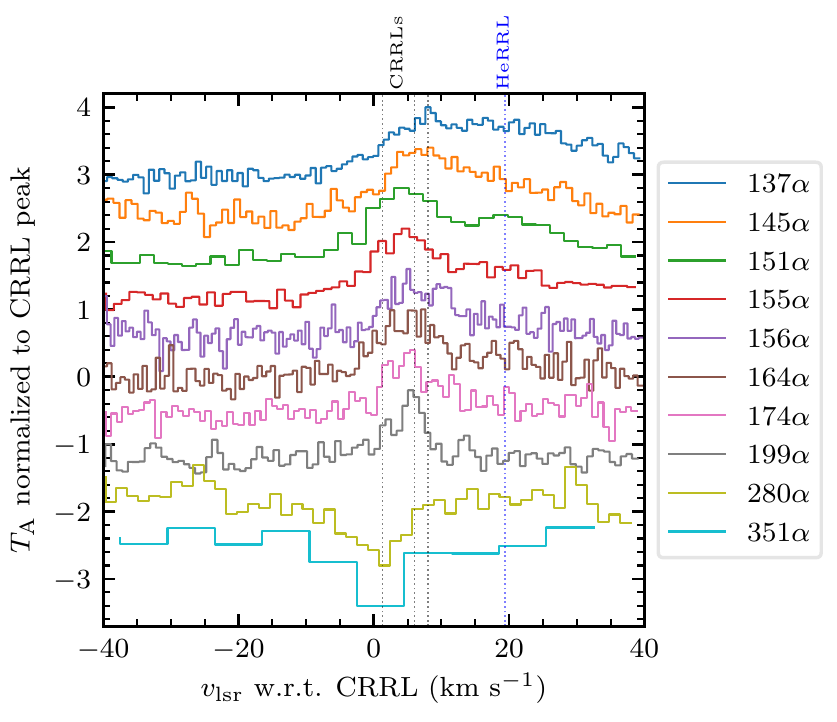}
 \caption{Zoom-in of RRL spectra towards M42 around the carbon feature. 
 The RRLs correspond to $\alpha$ lines with principal quantum numbers $137$, $145$, $155$, $156$, $164$, $174$, $199$, $280$, and $351$.
 CRRLs with $n\leq199$ appear in emission, while those with $n\geq280$ appear in absorption.
 The velocity is given with respect to the CRRL and the intensity axis is normalized to the peak of the CRRL.
 To reference the velocity with respect to helium,  $27.4$~km~s$^{-1}$ was subtracted.
 The spectra are offset by a constant $0.6$ and are normalized using the peak of the brightest CRRL in each spectra.
 The dotted lines indicate the position of the CRRLs at $\approx1.3$~km~s$^{-1}$, $\approx6$~km~s$^{-1}$, and $\approx8$~km~s$^{-1}$ (black) and the HeRRL (blue).
 The C$351\alpha$ spectrum is the spatial average over a circle $36\arcmin$ in diameter centered on M42.
 \label{fig:m42crrls}}
\end{figure}

Before focusing on the CRRLs we use the strength of the HRRLs to estimate the accuracy of the temperature scale: 
 HRRLs of similar principal quantum number have similar properties.
Then we can quantify the accuracy of the temperature scale by comparing the temperature of H$n\alpha$ lines of similar principal quantum number.
The peak temperatures of the HRRLs presented in Figure~\ref{fig:gbt2lbandcrrls} show variations of up to $25\%$ between adjacent stacks (see Table~\ref{tab:gbtcrrlsgfit}).
For example, the peak temperature of the H$155\alpha$ line should be almost the same as that of the H$156\alpha$ line, but they differ by $22\%$.
Based on this we conclude that the calibration using the noise diode has an accuracy of about $25\%$.

As pointed out above, the CRRLs can be identified as a narrow feature on top of the broader HeRRLs in Figure~\ref{fig:gbt2lbandcrrls}.
A zoom-in of the CRRLs is presented in Figure~\ref{fig:m42crrls}.
One of the most notable features in the spectra of Figure~\ref{fig:m42crrls} is the transition of the lines from emission to absorption between the C$199\alpha$ and C$280\alpha$ lines.
Towards M42, this is the first time that CRRLs have been observed in absorption.

In terms of the velocity structure of the CRRLs, we can identify at least two velocity components in emission at $6$ and $8$~km~s$^{-1}$.
The $\approx8$~km~s$^{-1}$ velocity component can be observed in the C$137\alpha$ RRL, while the $\approx6$~km~s$^{-1}$ velocity component can be observed in the CRRLs with $n=145\mbox{--}199$.
Gas with a velocity of $\approx8$~km~s$^{-1}$ is associated with the background molecular cloud, while gas with lower velocities is associated with foreground gas \citep[e.g.,][]{Dupree1974,Ahmad1976b,Boughton1978}.
In the case of this line of sight the foreground gas corresponds to the Veil, which is less dense \citep[$n_{\mathrm{H}}\sim10^{3}$~cm$^{-3}$][]{Abel2016} and irradiated by a weaker radiation field \citep[e.g.,][]{Abel2016} than the PDR that forms between the HII region and Orion~A \citep[$n_{\mathrm{H}}\sim10^{5}$~cm$^{-3}$; e.g.,][]{Natta1994}.

\begin{figure}
 \includegraphics[width=0.5\textwidth]{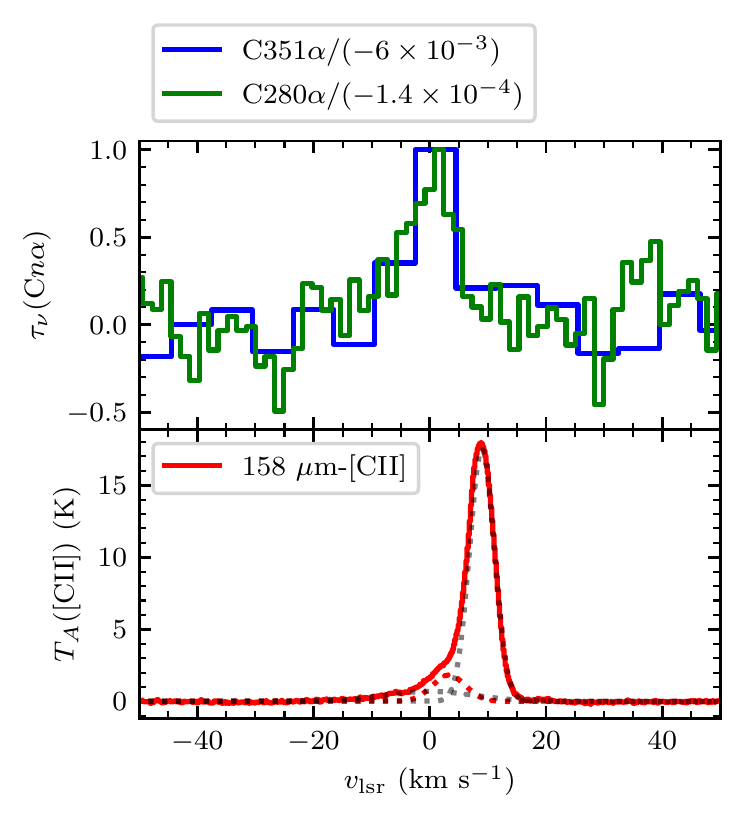}
 \caption{Comparison between the CRRLs observed in absorption and the $158$~$\mu$m-[CII] line.
           The blue steps show the C$351\alpha$ line profile inverted (from the LOFAR observations in $2014$), the green steps the C$280\alpha$ line inverted (from the GBT observations AGBT02A\_028), and the red steps show the $158$~$\mu$m-[CII] line (from the SOFIA observations  of \citealt{Pabst2019}).
          The CRRLs trace a fainter velocity component in the $158$~$\mu$m-[CII] line due to the effect of stimulated emission.
          The dotted lines in the lower panel show the best fit Gaussian line profiles used to decompose the $158$~$\mu$m-[CII] line (the properties of these components are given in Table~\ref{tab:veilfit}).
          The spectra are the spatial average over a circle $36\arcmin$ in diameter centered on M42.
          \label{fig:c351acii}}
\end{figure}

For the C$174\alpha$ and C$199\alpha$ lines there are hints of emission at $\approx2$~km~s$^{-1}$.
CRRL emission at this velocity has not been reported previously, though some authors reported the detection of unidentified RRLs at velocities of $\approx-3$~km~s$^{-1}$ \citep{Chaisson1974b} and $\approx-0.6$ \citep{Pedlar1974}.
Given that the $\approx2$~km~s$^{-1}$ velocity component is detected in two independent observations (the C$174\alpha$ stack is part of project AGBT02A\_028, while the C$199\alpha$ stack is part of AGBT12A\_484),  we consider the features to be CRRLs.
The C$174\alpha$ and C$199\alpha$ lines at $\approx2$~km~s$^{-1}$ trace gas in component B of the Veil.

To compare the lines in absorption we use an aperture of $36\arcmin$, similar to the resolution of the observations used to produce the C$280\alpha$ detection ($40\arcmin$, Table~\ref{tab:gbt2}).
The inverted spectra are presented in Figure~\ref{fig:c351acii}.
The C$280\alpha$ line has a velocity centroid of $0.7\pm1.0$~km~s$^{-1}$ (Table~\ref{tab:gbtcrrlsgfit}), while the C$351\alpha$ line has a velocity centroid of $2.3\pm0.8$~km~s$^{-1}$.
These lines trace the expanding Veil.

\begin{figure*}[!ht]
 \includegraphics[width=\textwidth]{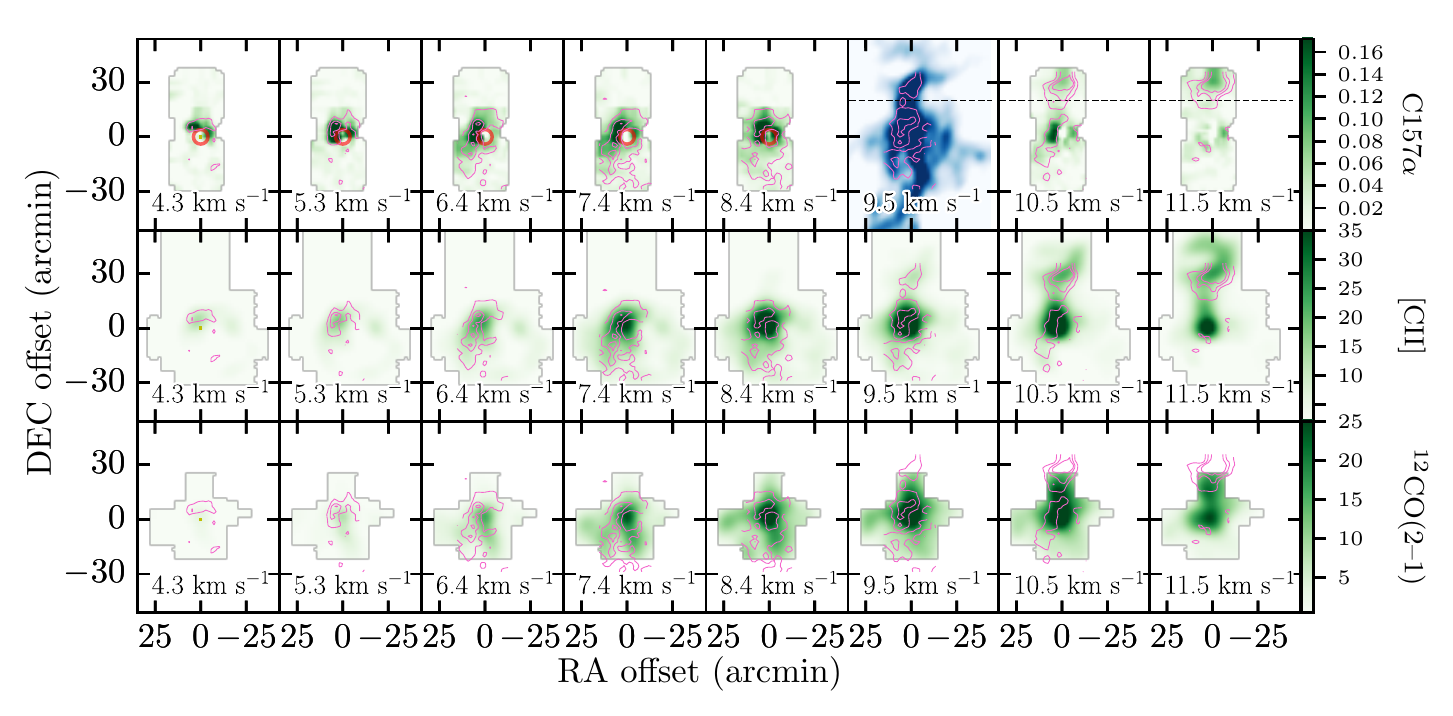}
 \caption{Channel maps of C$157\alpha$ (top row), $158$~$\mu$m-[CII] (middle row), and $^{12}$CO$(2\mbox{--}1)$ (bottom row) line emission.
          The pink contours show C$157\alpha$ emission above $3\sigma$ in steps of $3\sigma$, with $\sigma$ being the standard deviation of the spectra ($\sigma\approx 10$~mK).
          The velocity is indicated at the bottom of each panel.
          All cubes have been convolved to a spatial resolution of $8\farcm1$.
          The velocity axes were averaged to match the velocity resolution of the C$157\alpha$ cube.
          The spatial axes are given in offset with respect to M42.
          The red circle shows the extent of M42 in the $21$~cm continuum map of \citet{vanderWerf2013}.
          In the top panel with a velocity of $9.5$~km~s$^{-1}$ the background image in blue is the $857$~GHz emission as observed with Planck at $4\farcm6$ resolution \citep{PlanckI2016}.
          In the top row panels with a velocity $\geq9.5$~km~s$^{-1}$, the dashed line shows a Declination of $-5.0583\degr$ (J2000), used to separate S279 from Orion~A.
          The color scales at the right are in units of K.
          \label{fig:spatialcomp}}
\end{figure*}

The $158$~$\mu$m-[CII] line spectrum extracted from the $36\arcmin$ aperture used to study the C$280\alpha$ and C$351\alpha$ lines is also shown in Figure~\ref{fig:c351acii}.
There we see that the Veil ($v\approx3$~km~s$^{-1}$) has a peak antenna temperature of $\approx1.8$~K, while that from the background PDR ($v\approx9$~km~s$^{-1}$) is a factor of ten stronger.
The Veil is weaker in the $158$~$\mu$m-[CII] line because it is farther from the Trapezium \citep[$\approx2$~pc;][]{Abel2016} and hence colder.

\subsection{Spatial distribution of CRRLs}

\subsubsection{$\mathrm{C}157\alpha$}
\label{sssec:c157a}

The spatial distribution of a stack of CRRLs with $n=156\mbox{--}158$ (with an effective $n=157$) is presented in Figure~\ref{fig:spatialcomp} in the form of channel maps.
In Figure~\ref{fig:spatialcomp} we also include channel maps of $158$~$\mu$m-[CII] and $^{12}$CO$(2\mbox{--}1)$ at the same resolution.
The channel maps show that the C$157\alpha$ emission avoids the regions where the $18$~cm continuum is brightest.
At the frequency of the C$157\alpha$ line ($\approx1.6$~GHz) the brightest portions of the HII region are optically thick \citep[e.g.,][]{Wilson2015}.
This means that radiation coming from the interface between the background cloud and the HII region is heavily attenuated at these frequencies.
Moreover, the noise is greater towards the HII region due to its contribution to the antenna temperature.

At velocities of less than $6$~km~s$^{-1}$ the C$157\alpha$ emission comes from regions close to the Northern Dark Lane and the Dark Bay.
The Northern Dark Lane is a dark structure that separates M42 from M43 in optical images \citep[see Figure~12 in ][]{ODell2010}.
The Dark Bay is a region of high optical extinction which seems to start in the Northern Dark Lane and extends to the southwest in the direction of the Trapezium stars.
These structures are also seen in the lines of $158$~$\mu$m-[CII] and $^{12}$CO$(2\mbox{--}1)$.
At velocities in the range $6$~km~s$^{-1}$ to $7.4$~km~s$^{-1}$ the C$157\alpha$ emission extends to the  south of M42, following the limb brightened edge of the Veil \citep[][]{Pabst2019}.
Then at $8.4$~km~s$^{-1}$ the C$157\alpha$ emission seems to trace the Orion molecular cloud 4 \citep[OMC4, e.g.,][]{Berne2014}.
At velocities higher than $9$~km~s$^{-1}$ we see C$157\alpha$ emission extending to the north of M42.
At $10$~km~s$^{-1}$ we see part of the HII region S279 in the northernmost portion of the map (at an offset of $30\arcmin$ to the north), containing the reflection nebulae NGC 1973, 1975, and 1977.
In general, the spatial distribution of the C$157\alpha$ emission follows that of $158$~$\mu$m-[CII] and to a lesser extent that of $^{12}$CO$(2\mbox{--}1)$.
Then C$157\alpha$ emission predominantly traces the northern part of the ISF (see the top panel in Figure~\ref{fig:spatialcomp} for $v_{\mathrm{lsr}}=9.8$~km~s$^{-1}$).

\begin{figure}
 \includegraphics[width=0.5\textwidth]{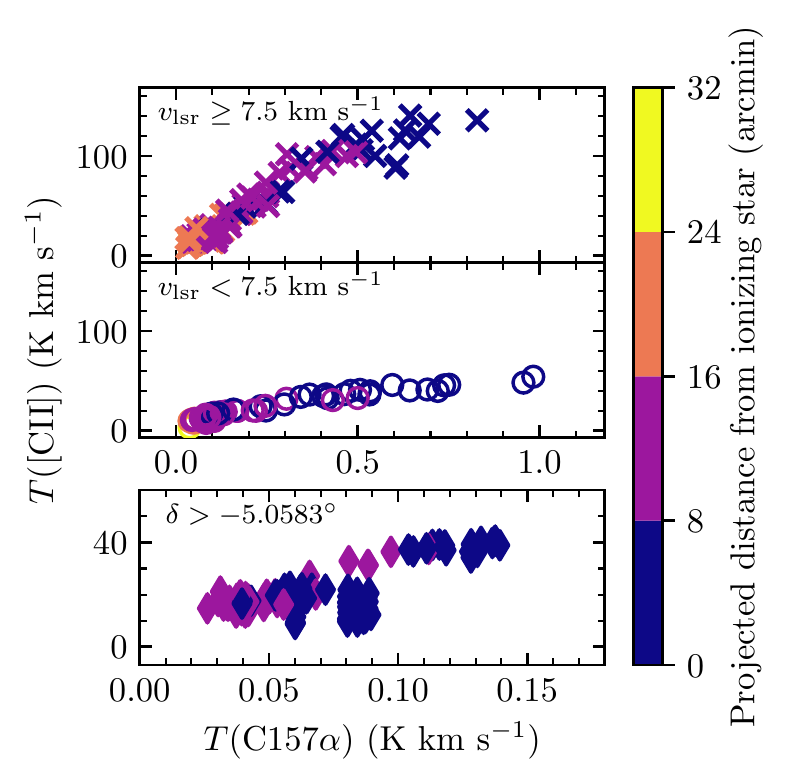}
 \caption{$158$~$\mu$m-[CII] line intensity as a function of the C$157\alpha$ line intensity.
          The line emission is separated into different groups based on known features in the maps.
          The top panel shows line emission with  velocity in the range $[4,7.5)$~km~s$^{-1}$ and  declination below $-5.0583\degr$, associated with the Veil; the middle panel shows line emission with  velocity in the range $[7.5,12)$~km~s$^{-1}$ and  declination below $-5.0583\degr$, associated with the ISF; and the bottom panel shows line emission at  declination above $-5.0583\degr$, associated with S279.
          \label{fig:cii_crrl}}
\end{figure}

To further explore the relation between the FIR [CII] line and the C$157\alpha$ line we compare their intensities at each position in the map.
We select pixels that show C$157\alpha$ emission with a signal-to-noise ratio $\geq5$ in the velocity range $4\mbox{--}12$~km~s$^{-1}$.
We split the selected pixels into three groups that separate different components in Orion~A.
The first group aims to trace gas along the ISF.
For this group we select pixels with line emission in the velocity range $7.5\leq v_{\mathrm{lsr}}<12$~km~s$^{-1}$ and with a declination below $-5.0583\degr$ (J2000).
The second group targets gas that is associated with the Veil.
Pixels with line emission in the velocity range $4\leq v_{\mathrm{lsr}}<7.5$~km~s$^{-1}$ and a declination below $-5.0583\degr$ (J2000) are selected in this group.
The third group targets gas associated with S279.
In this group, pixels with a declination above $-5.0583\degr$ (J2000) are selected.

The $158$~$\mu$m-[CII] and C$157\alpha$ line intensities for the different groups are presented in Figure~\ref{fig:cii_crrl}.
Here we can see that there is a relation between the intensities of both lines, and that the shape of their relation depends on which velocity structure is selected.
Gas associated with the ISF reaches a higher $158$~$\mu$m-[CII] line brightness than that in the other groups (the Veil or S279).
The shape of the relation for the gas associated with S279 looks like a scaled-down version of that in the ISF.
For the gas in the Veil, the C$157\alpha$ line is brighter than in the ISF or S279 at similar $158$~$\mu$m-[CII] brightness temperature, because the Veil is in front of the continuum source.

\begin{sloppypar}
In Figure~\ref{fig:cii_crrl} we have also color-coded the data as a function of their projected distance from the ionizing star.
For the gas in the ISF and the Veil, $\Theta^{1}$~Ori~C (HD~37022) is the ionizing star, while for gas in S279 it is $42$~Ori (HD~37018, c~Ori);
$\Theta^{1}$~Ori~C is a O7 star, while $42$~Ori is a B1 star \citep{Hoffleit1995}.
There is a trend in the line brightness as a function of distance from the ionizing star; closer to the ionizing source the lines are brighter.
\end{sloppypar}

In the CRRL spectra of Figure~\ref{fig:m42crrls} we can see that as the frequency decreases (increasing $n$), the velocity centroid of the emission lines shifts from $\approx9$~km~s$^{-1}$ to $\approx6$~km~s$^{-1}$ and additional velocity components are more easily observed at lower frequencies (e.g., at $2$~km~s$^{-1}$).
This is due to a combination of effects.
First, the dominant emission mechanism changes as a function of frequency.
At higher frequencies spontaneous emission dominates, while at lower frequencies stimulated transitions become dominant (Sect.~\ref{sssec:veilprops}).
Spontaneous emission lines are brighter from denser regions (i.e., the background PDR), while to get stimulated transitions a bright background continuum is required.
Second, all the observations were obtained using the same telescope, hence the observing beam becomes larger with decreasing frequency, and therefore different gas structures are included in the beam.
As Figure~\ref{fig:spatialcomp} shows, the velocity distribution of the gas is such that gas with lower velocities has a higher emission measure around M42 than towards M42 itself.
This implies that at higher frequencies we mainly see CRRLs from the background PDR since this is the densest component along the line of sight, while at lower frequencies we observe the gas around and in front of M42.

\subsubsection{$\mathrm{C}30\alpha$}

\begin{figure}
 \includegraphics[width=0.5\textwidth]{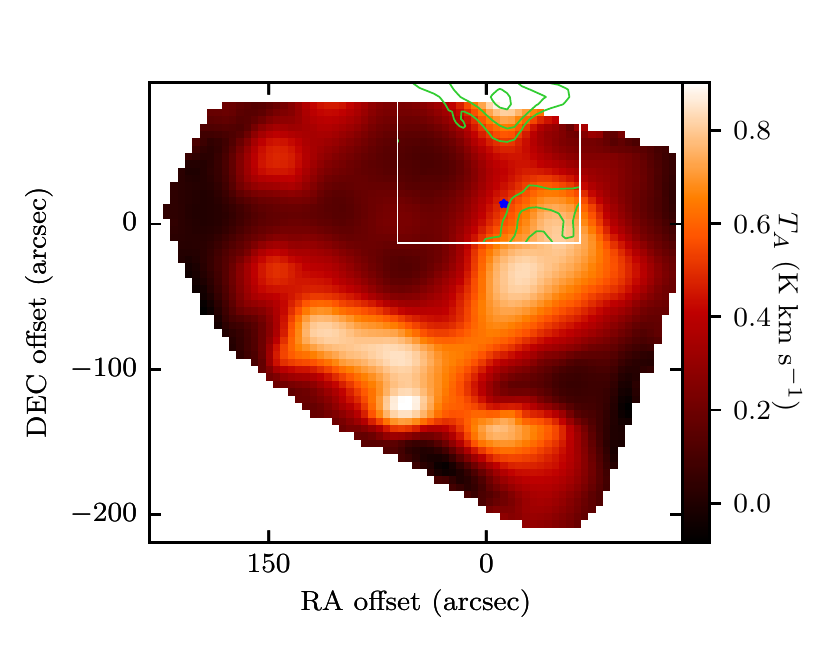}
 \caption{Moment $0$ maps of C$30\alpha$ emission and C$65\alpha$ emission.
          The green contours show the C$30\alpha$ emission at values of $40$, $60,$ and $80$~mK~km~s$^{-1}$.
          The color map shows the C$65\alpha$ emission \citep{Wyrowski1997}.
          The spatial resolution of the C$30\alpha$ map is $28\arcsec$, while that of the C$65\alpha$ map is $40\arcsec$.
          A white box shows the extent of the region mapped by ALMA where C$30\alpha$ is detected \citep[southeast map in ][]{Bally2017}.
          The spatial axes are given in offset with respect to M42, and a blue star indicates the position of $\Theta^{1}$~Ori~C.
          \label{fig:c30mom0}}
\end{figure}

We searched for CRRLs in the ALMA cubes presented by \citet{Bally2017}.
These cubes contain $\alpha$ RRLs with $n=30$ within the observed frequency range. The
H$30\alpha$, He$30\alpha$, and C$30\alpha$ lines are detected in the cube that covers the southeast region of the Orion Molecular Core 1.
We confirm that the observed line is C$30\alpha$ by comparing its velocity integrated intensity (moment $0$) with that of the C$65\alpha$ line at a similar angular resolution \citep[$40\arcsec$,][]{Wyrowski1997}.
The comparison is presented in Figure~\ref{fig:c30mom0}, where we can see the C$30\alpha$ emission overlapping with the C$65\alpha$ emission over the region mapped.
This confirms that the emission corresponds to C$30\alpha$ and not to a molecular line at a similar velocity.

\begin{figure}
 \includegraphics[width=0.5\textwidth]{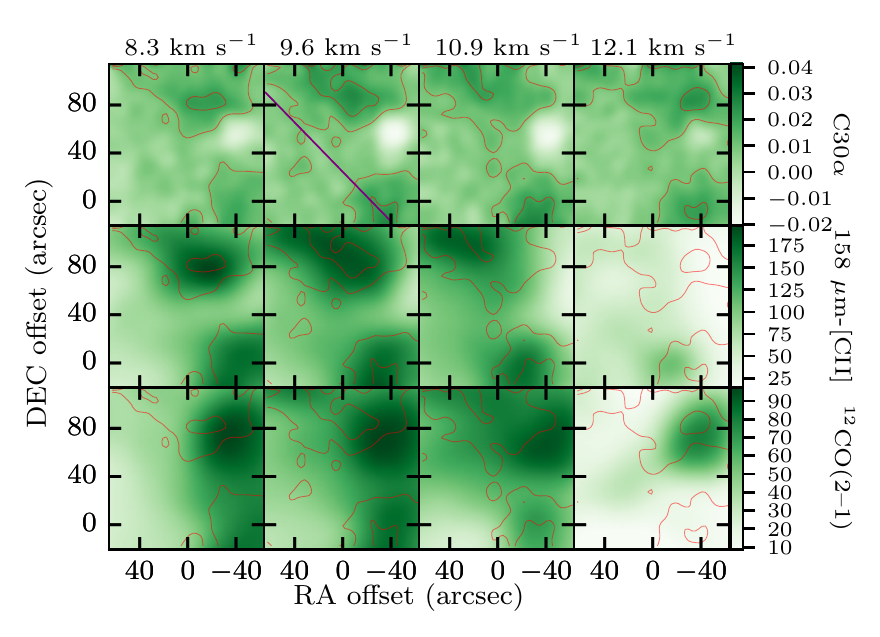}
 \caption{Channel maps of C$30\alpha$ (top row), $158$~$\mu$m-[CII] (middle row), and $^{12}$CO$(2\mbox{--}1)$ (bottom row) line emission.
          The red contours show C$30\alpha$ emission above $10$~mK, in steps of $10$~mK.
          The velocity with respect to the local standard of rest is indicated at the top of each row.
          All cubes have been convolved to a spatial resolution of $28\arcsec$.
          The velocity axes were averaged and then linearly interpolated to match the velocity axis of the C$30\alpha$ cube.
          The spatial axes are given in offset with respect to M42.
          In the C$30\alpha$ panel with a velocity of $9.6$~km~s$^{-1}$ a solid purple line shows the slice used to extract the brightness profile presented in Figure~\ref{fig:slices}.
          \label{fig:c30amap}}
\end{figure}

Next we examine how the C$30\alpha$ emission is distributed with respect to the $158$~$\mu$m-[CII] and $^{12}$CO$(2\mbox{--}1)$ lines.
This comparison is presented in Figure~\ref{fig:c30amap}.
The distribution of C$30\alpha$ resembles that of the other two lines, but there are differences between them.

\begin{figure}
 \includegraphics[width=0.5\textwidth]{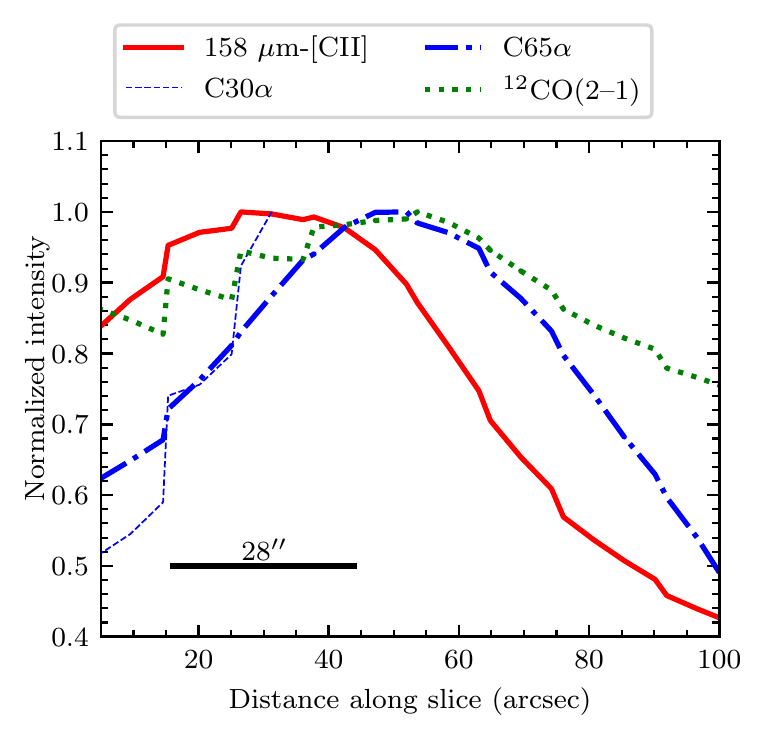}
 \caption{Comparison of the intensities of the C$30\alpha$, $158$~$\mu$m-[CII] \citep[][]{Pabst2019}, and $^{12}$CO$(2\mbox{--}1)$ \citep{Berne2014} lines.
          The thin dashed blue line shows the C$30\alpha$ line profile, the blue dash-dotted line for the C$65\alpha$ line, the red solid line for the $158$~$\mu$m-[CII] line, and the green dotted line for the $^{12}$CO$(2\mbox{--}1)$ line.
          The slice from where the velocity integrated brightness profiles was extracted is shown in Figure~\ref{fig:c30amap}.
          The position of $\Theta^{1}$~Ori~C marks the origin of the distance scale.
          \label{fig:slices}}
\end{figure}

To illustrate the above point we extract the line intensity from a slice that joins $\Theta^{1}$~Ori~C with the peak of C$30\alpha$ emission in the south of the map (purple line in Figure~\ref{fig:c30amap}).
To produce the intensity profiles the cubes are integrated over the velocity range $8$~km~s$^{-1}$ to $12$~km~s$^{-1}$, and the result is presented in Figure~\ref{fig:slices}.
There we can see that the $158$~$\mu$m-[CII] line peaks closer to $\Theta^{1}$~Ori~C than the $^{12}$CO$(2\mbox{--}1)$ line and the CRRLs.
This arrangement is similar to the layered structure found in a PDR \citep[e.g.,][]{Wyrowski2000}. 

\subsection{PDR models}
\label{ssec:pdrmod}

To understand the relation between the gas traced by the $158$~$\mu$m-[CII] line and that traced by the CRRLs we use a PDR model.
In this case we use the \emph{Meudon} PDR code \citep{Lepetit2006} to generate temperature and density profiles.
To model the PDR we adopt a total extinction of $A_{V}=20$ along the line of sight and a constant thermal pressure throughout the gas slab.
How far  the UV radiation penetrates into the PDR is largely determined by the extinction curve, which towards $\Theta^{1}$~C~Ori is almost flat with an extinction-to-color index $R_{V}=5.5$ \citep{Fitzpatrick1988,Cardelli1989}.
The extinction-to-column density ratio ($A_{V}/N_{\mathrm{H}}$) is determined from the extinction observed towards the Trapezium stars, $A_{V}=2.13\pm0.52$ \citep[][]{Ducati2003}, and the hydrogen column density towards $\Theta^{1}$ Ori C and B of $N_{\mathrm{H}}=4.4\times10^{21}$~cm$^{-2}$ \citep{Shuping1997,Cartledge2001}.
We adopt a carbon abundance of [C/H]$=1.4\times10^{-4}$, measured against $\Theta^{1}$~Ori~B \citep{Sofia2004}.
The models are illuminated by the ISRF on the far side ($A_{V}=20$) scaled to $G_{0}=1$ using the parametrization of \citet{Mathis1983}.
On the observer side ($A_{V}=0$) we vary the strength of the ISRF to explore its effect on the gas properties.

\begin{figure}
 \includegraphics[width=0.5\textwidth]{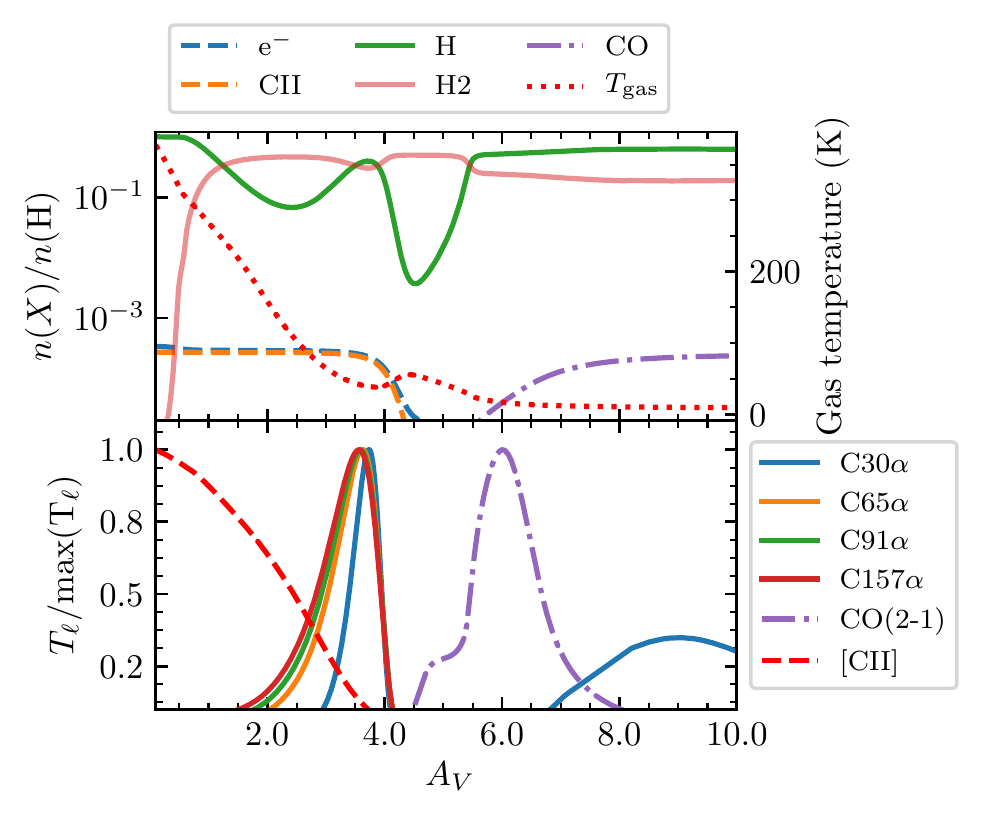}
 \caption{Example of temperature and abundance profiles obtained with the \emph{Meudon PDR code}.
          The top panel shows the gas temperature and abundances, while the bottom panel shows the line brightness temperature of C$n\alpha$ lines with principal quantum numbers $n=30,\,65,\,91\,\mbox{and}\,157$, and of the $158$~$\mu$m-[CII] line.
          The input conditions for the model are a radiation field of $G_{0}=1\times10^{4}$, in Mathis units, and a total gas density of $n_{\mathrm{H}}=1\times10^{4}$~cm$^{-3}$.
          The difference between the abundance of free electrons and the abundance of ionized carbon is produced by the ionization of species such as sulfur or hydrogen.
          \label{fig:pdr}}
\end{figure}

Once we have computed the temperature and density in the PDR, we process the output to determine how much of the $158$~$\mu$m-[CII] and CRRL brightness comes from different layers in the PDR.
The different layers represent different depths into the molecular cloud and are expressed in terms of the visual extinction $A_{V}$.
Examples of the temperature and density profiles, and of  the line brightness contributed from each layer in the PDR, are presented in Figure~\ref{fig:pdr}.
For this model we used an incident radiation field of $G_{0}=1\times10^{4}$, in Mathis units, and a total gas density of $n_{\mathrm{H}}=1\times10^{4}$~cm$^{-3}$.
The layered structure in the models is in good agreement with observations of CRRLs and the $158$~$\mu$m-[CII] line for the PDRs associated with the Orion Bar and NGC 2023 \citep[][]{Wyrowski1997,Wyrowski2000,Bernard-Salas2012,Sandell2015}.

We use the models of \citet{Salgado2017a} to compute the properties of the CRRLs.
These models solve the level population equations taking into account deviations from local thermodynamical equilibrium (LTE).
The deviation from LTE in the population of carbon atoms is characterized by the factor $b_{n}$ and the effect of stimulated emission by the factor $\beta_{nn^{\prime}}$ \citep[e.g.,][]{Shaver1975,Salgado2017a}.
These are known as departure coefficients.
The models of \citet{Salgado2017a} include the effect of dielectronic capture \citep{Watson1980,Walmsley1982}.
This effect will produce an overpopulation at $n$ levels in the range $30$--$500$ with respect to a system that does not undergo dielectronic capture.
For conditions like those found towards Orion~A \citep[$n_{\mathrm{H}}\sim10^{5}$ and $T\sim100$~K, e.g.,][]{Natta1994}, dielectronic capture will produce twice as many atoms with an electron at $n=91$ than if we ignore its effect.
The effect of dielectronic capture has not been considered explicitly before when studying the Orion~A region, but it has been suggested that it could help explain the observed line ratios \citep[][]{Wyrowski1997}.
We note that when solving the level population problem we do not include the presence of a free-free radiation field. 
For hydrogen atoms, the departure coefficients will change by less than $12\%$ for $n$ between $10$ and $60$ \citep[e.g.,][]{Prozesky2018}.
The effect is smaller for $n>60$.

For a homogeneous slab of gas in front of a continuum source, the intensity of a CRRL, $T_{\ell}\Delta\nu$, is given by \citep[e.g.,][]{Dupree1974}
\begin{equation}
 T_{\ell}\Delta\nu=\tau^{*}_{\ell}\Delta\nu(b_{n^{\prime}}T_{\mathrm{e}}-b_{n}\beta_{nn^{\prime}}T_{\mathrm{cont}}),
 \label{eq:crrl_ta}
\end{equation}
where $\tau^{*}_{\ell}$ is the line optical depth in LTE, $T_{\mathrm{e}}$ the electron temperature of the gas, and $T_{\mathrm{cont}}$ the temperature of the background continuum.
In this equation, the first term in parentheses corresponds to the contribution to the line brightness temperature from spontaneous emission, while the second term represents the contribution from stimulated emission.
The line optical depth in LTE is given by \citep[e.g.,][]{Salgado2017b}
\begin{equation}
 \tau^{*}_{\ell}\Delta\nu=1.069\times10^{7}\Delta nMT_{\mathrm{e}}^{-2.5}e^{\chi_{n}}EM_{\mathrm{C}^{+}}\,\mbox{Hz},
 \label{eq:crrl_tau}
\end{equation}
where $\Delta n=n^{\prime}-n$, $M$ is the oscillator strength of the transition \citep{Menzel1968}, $\chi_{n}=157800n^{-2}T^{-1}_{\mathrm{e}}$, and $EM_{\mathrm{C}^{+}}=n_{\mathrm{e}}n_{\mathrm{C}^{+}}L$ is the ionized carbon emission measure in pc~cm$^{-6}$ with $L$ the thickness of the slab.

To compute the CRRL brightness temperature from the PDR we assume that the emission is due to spontaneous emission with no background continuum \citep{Natta1994}.
In each layer the temperature and electron density determine the value of $b_{n}$.
For the CRRLs the $b_{n}$ values are $<1$ over the range of physical properties explored here.
The line brightness in the LTE case is, on average,  $50\%$ larger than in the non-LTE case.
The difference between the LTE and non-LTE cases is larger for lower pressures and higher radiation fields.
In an extreme case the LTE value is $70\%$ higher than the non-LTE value.

To compute the $158$~$\mu$m-[CII] line brightness temperature we use the equations provided in Appendix B of \citet{Tielens1985a} and the collisional excitation rates provided in \citet{Goldsmith2012}.
The equations in \citet{Tielens1985a} provide the line intensity with a correction for the finite optical depth of the line.

In Figure~\ref{fig:pdr} we can see that most of the $158$~$\mu$m-[CII] line comes from the surface layers of the PDR ($A_{V}<3.5$), while the CRRL emission comes from a deeper layer ($A_{V}=3.5$).
The gas temperature can be a factor of $10$ lower at $A_{V}=3.5$ with respect to $A_{V}<3.5$.
This shows that the CRRL optical depth has a stronger dependence on the temperature ($\propto T^{-5/2}$) than that of the $158$~$\mu$m-[CII] line.
Therefore, when we constrain the gas physical properties using CRRLs and the $158$~$\mu$m-[CII] line using a uniform gas slab model, the temperature and density will be an average between the properties of the layers traced by the two lines.
We also note that the studied CRRLs trace an almost identical layer in the PDR, which justifies using their line ratios regardless of geometry.
The situation is similar for PDRs with $10^{2}<G_{0}\leq10^{5}$ and $10^{2}$~cm$^{-3}<n_{\mathrm{H}}\leq10^{6}$~cm$^{-3}$.

The structure observed in Figure~\ref{fig:pdr} is similar to that found in Figure~\ref{fig:slices}.
There, we observe that the separation between the peak of the $158$~$\mu$m-[CII] line is offset by $\approx10\arcsec$ with respect to the peak of $^{12}$CO$(2\mbox{--}1)$.
For a distance of $417$~pc, this translates to a projected separation of $0.02$~pc.
Using the result of Figure~\ref{fig:pdr}, we have that the separation between these tracers corresponds to roughly $A_{\mathrm{V}}=6$ or $N_{\mathrm{H}}=1.2\times10^{22}$~cm$^{-2}$.
This corresponds to a hydrogen density of $2\times10^{5}$~cm$^{-3}$, similar to that found in the interclump medium in the Orion Bar ($5\times10^{4}$~cm$^{-3}$ \citealt{YoungOwl2000} or $2\times10^{5}$~cm$^{-3}$ \citealt{Simon1997}).
This hydrogen density is also consistent with the value found towards a nearby region using CRRL and [CII] ratios (Sect.~\ref{sssec:pdrgas}).

\section{Physical conditions}
\label{sec:physcond}

In this section we use CRRLs and the $158$~$\mu$m-[CII] line to determine the physical conditions of the gas, for example its temperature and density.
We do this by modeling the change in the properties of the CRRLs as a function of principal quantum number \citep[e.g.,][]{Ahmad1976b,Boughton1978,Jaffe1978,Payne1994,Oonk2017,Salas2018}, and by comparing the CRRLs with different principal quantum numbers to the $158$~$\mu$m-[CII] line.

\subsection{The Veil towards M42}

To study the Veil of Orion we focus on the information provided by the CRRLs observed in absorption and the C$157\alpha$ emission at velocities $\lesssim7$~km~s$^{-1}$.

\subsubsection{Transition from emission to absorption}

For the CRRLs associated with the Veil the largest principal quantum number for which the line is observed in emission is $n=199$ (Figure~\ref{fig:m42crrls}).
Then at $n=280$ the line is observed in absorption.
This sets a lower limit to the electron density of the gas of $n_{\mathrm{e}}\geq0.03$~cm$^{-3}$, and for the electron temperature $35~\mbox{K}\leq T_{\mathrm{e}}\leq130$~K.
The constraint on the gas properties set by the transition from emission to absorption is shown in Figure~\ref{fig:veilprops} as a purple dashed line.

\subsubsection{CRRL ratio}
\label{sssec:crrls}

The ratio of two CRRLs in absorption provides an additional constraint to determine the gas properties \citep[e.g.,][]{Salgado2017b,Salas2017,Salas2018}.
Here we use the ratio of the integrated optical depths of the C$280\alpha$ and C$351\alpha$ lines to constrain the gas temperature and electron density.

In order to convert the observed C$280\alpha$ line temperature to optical depth, we need to estimate the continuum adjacent to the line.
As mentioned in Section~\ref{sssec:gbtm42}, we chose not to directly estimate the continuum from the observations used to produce the C$280\alpha$ spectrum as we do not have a reference position to use  to estimate the contribution from non-astronomical sources to the antenna temperature.
Instead, we use the low frequency spectrum of M42 to estimate the contribution to the continuum in the C$280\alpha$ spectrum.
Using the Very Large Array \citep[VLA,][]{Napier1983} in its D configuration (minimum baseline $35$~m), \citet{Subrahmanyan2001} observed M42 at $330$~MHz.
They measured a total combined flux for M42 and M43 (which is only $\sim5\arcmin$ away from M42) of $167\pm5$~Jy, consistent with single dish measurements \citep[e.g.,][]{Lockman1975}.
We assumed that the combined flux density from M42 and M43 scales as $S_{\nu}\propto\nu^{0.92\pm0.08}$ between $240$ and $400$~MHz \citep[based on the continuum measurements presented in][]{Lockman1975}.
We estimated the effect of beam dilution on the measured antenna temperature for the continuum using the $330$~MHz continuum maps \citep{Subrahmanyan2001}.
In these maps, M42 and M43 cover a circular area with a radius of $18\arcmin$ centered at $(\alpha,\delta)_{\mathrm{J2000}}=(5^{\mathrm{h}}35^{\mathrm{m}}00^{\mathrm{s}},-5\degr25^{\mathrm{m}}22^{\mathrm{s}})$.
The $330$~MHz continuum shows a structure  similar to that of the LOFAR $149$~MHz continuum map.
The beam of the C$280\alpha$ observations covers most of this region, and leaves out less than $0.4\%$ of the continuum flux.
Therefore, we estimated that at $298$~MHz the continuum temperature of the C$280\alpha$ spectra will be $195\pm6$~K.
Ultimately, we find the integrated optical depth of the C$280\alpha$ line is $1.4\pm0.2$~Hz.

\begin{table}
\caption{Veil line properties}
\begin{tabular}{lccc}
\toprule
Line   & $v_{\mathrm{lsr}}$ & $T_{\mathrm{line}}$ & $\Delta v$(FWHM) \\
       & (km s$^{-1}$)      & (K)                 & (km s$^{-1}$)    \\
\midrule
\multirow{3}{*}{[CII]} &  $8.98\pm0.01$ & $17.58\pm0.07$                      &  $5.02\pm0.01$ \\
                       &  $3.2\pm0.1$   &  $1.82\pm0.05$                      &  $6.6\pm0.2$   \\
                       & $-0.1\pm0.4$   &  $0.70\pm0.04$                      & $17.9\pm0.5$   \\
C$280\alpha$           &  $0.7\pm1.0$   & $-0.023\pm0.003$\tablefootmark{a}   & $11\pm1$       \\
C$351\alpha$           &  $2.3\pm0.8$   & $-0.0061\pm0.0008$\tablefootmark{b} & $10\pm1$       \\
\bottomrule                                         
\end{tabular}
\tablefoot{
The line properties correspond to the best fit parameters of Gaussian line profiles and the errors quoted are 1 $\sigma$. The fits were performed to the spectra presented in Figure~\ref{fig:c351acii}.\\
\tablefoottext{a}{To convert to optical depth we adopted a continuum temperature of $195\pm6$~K.}
\tablefoottext{b}{Optical depth. The flux density of Orion~A and M43 measured from the LOFAR continuum image at $149$~MHz is $53\pm3$~Jy.}
}
\label{tab:veilfit}
\end{table}

\begin{figure}
 \includegraphics[width=0.5\textwidth]{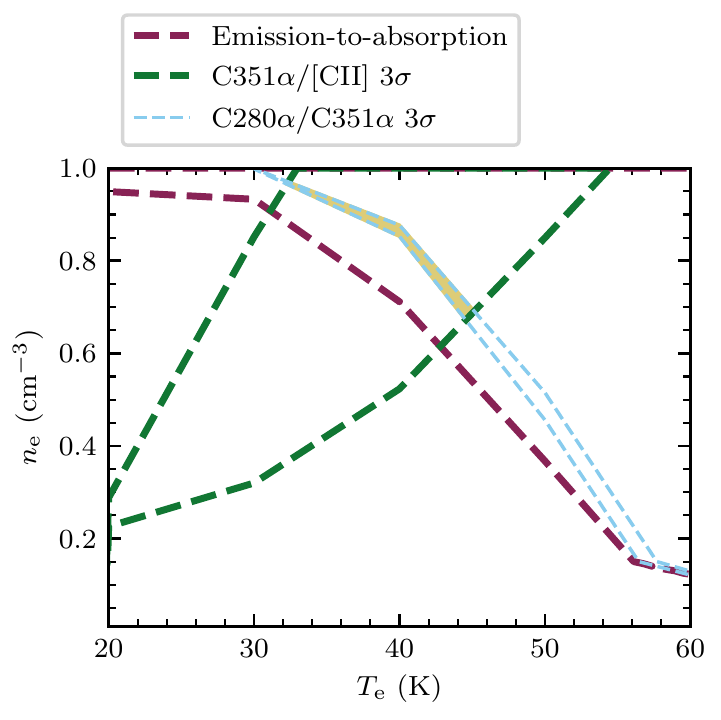}
 \caption{Constraints on the temperature and electron density for gas associated with Orion's Veil.
          The dashed lines show the constraints on the gas properties derived from different observables: the transition of the CRRLs from emission to absorption between $n=200$ and $279$ (purple); the ratio of the integrated optical depths of the C$280\alpha$ and C$351\alpha$ lines (light blue); the ratio of the C$351\alpha$ velocity integrated optical depth to the $158$~$\mu$m-[CII] line intensity (green).
          All the constraints shown are $3\sigma$ ranges.
          The region where the constraints overlap is shown as a yellow shaded region, close to $0.9$~cm$^{-3}$ and $40$~K.
          \label{fig:veilprops}}
\end{figure}

The ratio of the integrated optical depths of the C$280\alpha$ and C$351\alpha$ lines is $(4\pm1)\times10^{-2}$.
The constraint on the gas temperature and electron density set by this ratio is shown in Figure~\ref{fig:veilprops} with blue dashed lines.
A higher ratio implies a higher temperature.
In this case, the integrated optical depth ratio poses a more stringent constraint on the gas properties than the point at which the lines transition from emission to absorption.

\subsubsection{CRRLs and FIR [CII] line}

Here we compare the latest $158$~$\mu$m-[CII] line maps of \citet[][]{Pabst2019}  with the CRRLs observed in absorption.
The cube of \citet[][]{Pabst2019} presents the $158$~$\mu$m-[CII] line resolved in velocity and samples a region larger than that studied in CRRLs.
With this we were able to  perform a direct comparison between the lines over the same regions without making assumptions about their velocity structure.
Previous comparisons between CRRLs and the $158$~$\mu$m-[CII] line were performed using observations that did not resolve the velocity structure and/or did not sample the same spatial regions \citep[e.g.,][]{Natta1994,Smirnov1995,Salas2017}.

Here, we compare the C$351\alpha$ line with the $158$~$\mu$m-[CII] line over the same spatial regions.
Since the C$351\alpha$ line is observed in absorption, it will only trace gas that is in front of the continuum source.
Then the $158$~$\mu$m-[CII] line spectrum used to make a  comparison with the C$351\alpha$ line should be extracted from a region that encompasses the continuum source.
This corresponds to a circular region with a radius of $18\arcmin$ centered at $(\alpha,\delta)_{\mathrm{J2000}}=(5^{\mathrm{h}}35^{\mathrm{m}}00^{\mathrm{s}},-5\degr25^{\mathrm{m}}22^{\mathrm{s}})$.
The absorption spectra will be weighted by the underlying continuum, whereas the $158$~$\mu$m-[CII] line will not be.
Hence, even if we use an aperture that covers most of the continuum emission, the lines could trace different portions of the Veil.

The resulting $158$~$\mu$m-[CII] line spectrum (Figure~\ref{fig:c351acii}) shows the presence of at least three velocity components.
We fitted three Gaussian components corresponding to the Veil, the dense PDR, and the HII region.
The best fit parameters of the Gaussian profiles are presented in Table~\ref{tab:veilfit}.
Using the values for the component associated with the Veil, at $\approx3$~km~s$^{-1}$ the ratio of the C$351\alpha$ line integrated optical depth to the $158$~$\mu$m-[CII] line intensity is $(-378\pm76)\times10^{3}$~Hz~erg$^{-1}$~s~cm$^{2}$~sr$^{1}$.

Given that the brightness of the $158$~$\mu$m-[CII] line is $1.82\pm0.05$~K, and that the hydrogen density in the Veil is $\approx10^{3}$~cm$^{-3}$ \citep[][]{Abel2016}, we assume that the line is effectively optically thin \citep[EOT,][]{Goldsmith2012}.
In this case the intensity of the $158$~$\mu$m-[CII] line is proportional to the column density, hence the ratio with respect to the integrated optical depth of the C$351\alpha$ line is independent of the column density and the line width.
The constraints imposed on the gas properties based on the ratio of the $158$~$\mu$m-[CII] line intensity to the C$351\alpha$ line integrated optical depth are shown in Figure~\ref{fig:veilprops} with green dashed lines.

\subsubsection{Combined constraints: gas temperature and density}
\label{sssec:veilprops}

The constraints imposed on the gas properties by the integrated optical depth of the C$280\alpha$ and C$351\alpha$ lines and the ratio of the integrated optical depth of the C$351\alpha$ line to the $158$~$\mu$m-[CII] line intensity intersect (see Figure~\ref{fig:veilprops}).
The region where these constraints intersect determines the ranges of temperature and electron density allowed by our analysis.
The range of physical properties is then $30$~K$\leq T_{\mathrm{e}}\leq45$~K and $0.65$~cm$^{-3}\leq n_{\mathrm{e}}\leq0.95$~cm$^{-3}$ if we consider the $3\sigma$ ranges.
These constraints are valid for the Veil at $\approx3$~km~s$^{-1}$, under the assumption that the C$280\alpha$, C$351\alpha$, and $158$~$\mu$m-[CII] lines trace the same gas.
This assumption is appropriate for gas exposed to a radiation field $G_{0}\lesssim10^{3}$, when the temperature difference between the layers traced by the CRRLs and the $158$~$\mu$m-[CII] line is lower.
Since the gas properties were derived from line ratios, they do not have a strong dependence on the beam filling factor.

Using the derived gas properties and the observed brightness of the $158$~$\mu$m-[CII] line we can compute the column density of ionized carbon.
The intensity of the $158$~$\mu$m-[CII] line is $12.7\pm0.5$~K~km~s$^{-1}$ over a circular region with a $18\arcmin$ radius.
This implies that the beam averaged column density is $N_{\mathrm{CII}}=(3.0\pm0.4)\times10^{17}$~cm$^{-2}$, where the quoted $1\sigma$ error considers the $3\sigma$ range of possible physical properties.

\begin{figure}
 \includegraphics[width=0.5\textwidth]{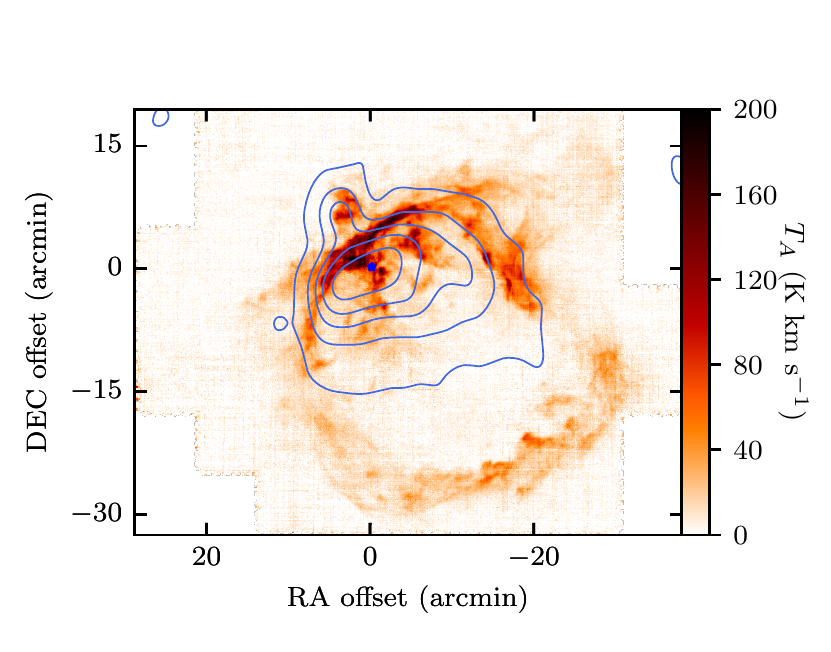}
 \caption{Moment 0 map of the $158$~$\mu$m-[CII] line associated with the Veil (color scale).
          The moment 0 map considers emission for velocities between $0$~km~s$^{-1}$ and $7$~km~s$^{-1}$.
          The contours show the radio continuum as observed with LOFAR at $149$~MHz.
          The contours start at $0.2$~mJy~beam$^{-1}$ and increase in steps of $1$~Jy~beam$^{-1}$.
          The spatial axes are given in offset with respect to M42, and a blue star indicates  the position of $\Theta^{1}$~Ori~C.
          The radio continuum partially fills the wind blown bubble.
          \label{fig:cont}}
\end{figure}

A closer inspection at the $158$~$\mu$m-[CII] line cubes at their native spatial resolution of $16\arcsec$ reveals that most of the emission at $v_{\mathrm{lsr}}\approx3$~km~s$^{-1}$ comes from the Dark Bay, the northern streamer \citep[see, e.g.,][]{Goicoechea2015}, part of M43, and the limb brightened Veil (\citealt{Pabst2019}, Figure~\ref{fig:cont}).
These cover an area of roughly $20\arcmin\times5\arcmin$ (Dark Bay plus northern streamer), $3\farcm5\times3\farcm5$ (M43) and $10\arcmin\times8\arcmin$ (limb brightened Veil) on the sky.
If we correct the column density for the effect of beam dilution we arrive at a value of $(2.3\pm0.4)\times10^{18}$~cm$^{-2}$, between the value towards the Dark Bay \citep[$1.5\times10^{18}$~cm$^{-2}$;][]{Goicoechea2015} and the limb brightened Veil ($3.5\times10^{18}$; \citealt{Pabst2019}).

We used the physical conditions we found to predict the peak antenna temperature of the C$157\alpha$ line.
We adopted the $3\sigma$ ranges for the gas properties, a full width at half maximum of $6$~km~s$^{-1}$, a column density of [CII] of $N_{\mathrm{CII}}=(3\pm0.4)\times10^{17}$~cm$^{-2}$, and a continuum temperature of $38$~K at $1.68$~GHz (over the $36\arcmin$ aperture).
The predicted line profile has a peak antenna temperature between $25$~mK and $170$~mK, consistent with the observed value of $70$~mK.
The range of predicted values is mainly determined by the gas temperature and density.
A variation of a factor of $1.5$ in density and in temperature translates to a factor of seven variation in antenna temperature because the departure coefficient $b_{n}\beta_{nn^{\prime}}$ is $20\%$ smaller in the high density--low temperature limit, but the exponential factor in the line optical depth (Equation~\ref{eq:crrl_tau}) is a factor of three larger, and the emission measure a factor of three larger.

For a gas temperature between $30$~K$\leq T_{\mathrm{e}}\leq45$~K and an electron density $0.65$~cm$^{-3}\leq n_{\mathrm{e}}\leq0.95$~cm$^{-3}$, the contribution to the antenna temperature due to spontaneous emission is $23\%$--$16\%$.
This implies that most of the C$157\alpha$ line emission associated with the Veil can be explained in terms of stimulated emission.
This reflects the importance of stimulated emission at low densities \citep[e.g.,][]{Shaver1975}.
For this range of physical conditions, the effects of spontaneous and stimulated emission become comparable at $n\approx120$.

The Veil has also been studied using other absorption lines: $21$~cm-HI, $18$~cm-OH, and lines in the ultraviolet (UV) \citep[e.g.,][]{vanderWerf1989,Abel2004,Abel2006,vanderWerf2013,Abel2016,Troland2016}.
Using observations of lines in the UV and the $21$~cm-HI line, \citet{Abel2016} have derived gas properties for components A and B of the Veil.
Their observations only sample the line of sight towards $\Theta^{1}$~Ori~C.
They find a gas density of $n_{\mathrm{H}}\approx10^{2.3}$ and $10^{3.4}$~cm$^{-3}$, and a temperature of $T_{\mathrm{K}}\approx50$ and $60$~K for components A and B, respectively.
Here we  used lower spatial resolution data to provide an average of the gas properties of the Veil in front of the HII region.
We find temperatures that are  $15\%$ lower than in the work of \citet{Abel2016}, which might mean that CRRLs trace lower temperature regions in a PDR.
To compare the density we need to convert from  electron density to  hydrogen density.
We assume that all of the electrons come from ionized carbon, $n_{\mathrm{e}}=n_{\mathrm{C}^{+}}$, and that the carbon abundance relative to hydrogen is $1.4\times10^{-4}$ \citep{Sofia2004}.
Then, our constraints on the electron density translate to  a hydrogen density $4000$~cm$^{-3}\leq n_{\mathrm{H}}\leq7000$~cm$^{-3}$, comparable to the values found by \citet{Abel2016}.
As the lack of C$137\alpha$ and C$145\alpha$ emission suggests, we do not expect the physical conditions to be uniform across the Veil.
This is confirmed by the patchy structure observed in $21$~cm-HI absorption \citep[][]{vanderWerf1989} and in optical extinction maps \citep[][]{ODell2000}.
Higher resolution observations of the C$280\alpha$ lines, or similar $n$ level, would allow us to study the temperature and density variations across the Veil.

\subsubsection{[CII] gas cooling and heating efficiency}

We estimate the gas cooling rate per hydrogen atom from the observed $158$~$\mu$m-[CII] intensity and the column density of hydrogen.
We convert the [CII] column density to a hydrogen column density assuming an abundance of carbon relative to hydrogen of $[\mathrm{C}/\mathrm{H}]=1.4\times10^{-4}$ \citep{Sofia2004} and that all carbon is ionized.
Under these assumptions, the observed intensity of the $158$~$\mu$m-[CII] line implies a [CII] cooling rate per hydrogen atom of $(4\pm0.2)\times10^{-26}$~erg~s$^{-1}$~(H-atom)$^{-1}$.
This is similar to the cooling rate found through UV absorption studies towards diffuse clouds \citep{Pottasch1979,Gry1992}; however, the Veil is exposed to a radiation field $\sim100$ higher than the average ISRF.
Given the geometry of the Veil, a large fraction of the $158$~$\mu$m-[CII] emission comes from regions that are optically thick towards the observer (\citealt{Pabst2019}, Figure~\ref{fig:cont}). 
Thus, the cooling rate we derive is likely a lower limit.

\begin{figure}
 \includegraphics[width=0.5\textwidth]{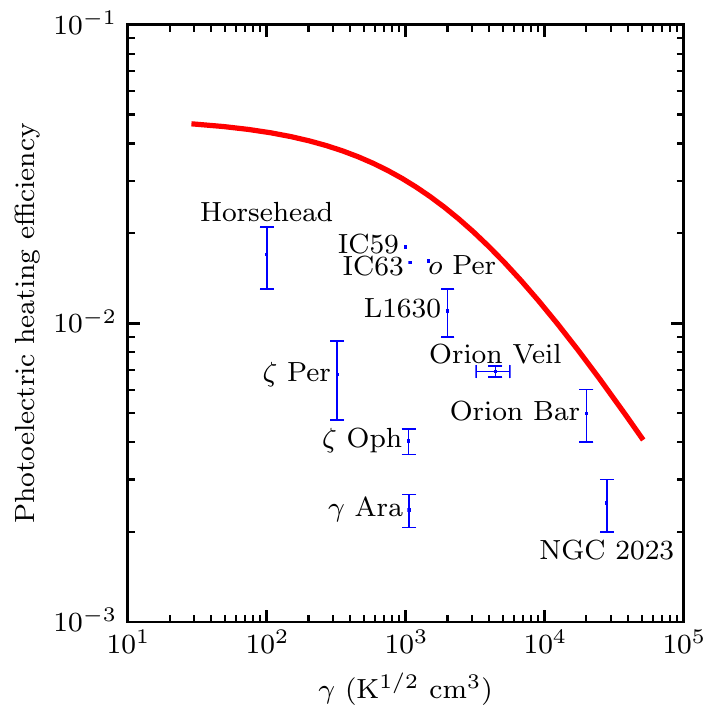}
 \caption{Photoelectric heating efficiency as a function of the ionization parameter $\gamma$.
          The data for the dense PDRs NGC 2023 and the Orion Bar is taken from \citet{Hollenbach1999}, the data for the Horsehead and L1630 is from \citet{Pabst2017}, the data for diffuse PDRs is from \citet{Gry1992} and \citet{vanDishoeck1986}, and the data for IC59 and IC63 is from \citet{Andrews2018}.
          The red line shows the model of \citet{Bakes1994}.
          The error bars for $o$~Per, IC59 and IC63 have been omitted for clarity.
          \label{fig:heat}}
\end{figure}

In the diffuse ISM most of the gas heating is through the photoelectric effect on polycyclic aromatic hydrocarbons (PAHs) and small dust grains \citep[e.g.,][]{Wolfire1995}.
In this process,  FUV ($6$~eV to $13.6$~eV) photons are absorbed by PAHs and very small dust grains causing them to eject electrons which then heat the gas through collisions.
Our understanding of the ISM is intimately related to the efficiency of this process, as it couples the interstellar radiation field to the gas temperature.
In general, the gas photoelectric heating efficiency $(\epsilon_{\mathrm{pe}}$) is less than $10\%$ \citep[e.g.,][]{Bakes1994,Weingartner2001a} and most of the energy absorbed by the dust is re-radiated in the infrared (IR).
Its exact value will depend on the charge state of the dust grains, and hence on the ionization parameter $\gamma=G_{0}T_{\mathrm{e}}^{1/2}n_{\mathrm{e}}^{-1}$ \citep[e.g.,][]{Hollenbach1999}.
The gas heating efficiency through the photoelectric effect can be estimated as ([CII]+[OI])/TIR \citep[e.g.,][]{Pabst2017}, where TIR is the total infrared flux and [OI] is the gas cooling through the line of atomic oxygen at $63$~$\mu$m.
Here we ignore the possible contribution from the [OI] line at $63$~$\mu$m to the gas cooling since for a gas density of $n_{\mathrm{H}}\approx3\times10^{3}$~cm$^{-3}$ it is estimated to be roughly $5\%$ of the total gas cooling \citep[e.g.,][]{Tielens2010}.
As a proxy for TIR we use the \citet{Lombardi2014} maps of dust properties.
These present the properties of the dust spectral energy distribution derived from fitting a modified blackbody to continuum data in the wavelength range $100$~$\mu$m to $3000$~$\mu$m as observed by Herschel and Planck.
From the maps of \citet{Lombardi2014} we can obtain the TIR flux by integrating the modified blackbody between the wavelength range $20$~$\mu$m to $1000$~$\mu$m. 
The median of the TIR flux over the $18\arcmin$ circle that contains the low-frequency radio continuum is $0.096$~erg~s$^{-1}$~cm$^{-2}$~sr$^{-1}$.
Then, if we correct for beam dilution, we have $\epsilon_{\mathrm{pe}}=(6.9\pm0.3)\times10^{-3}$.
For $G_{0}$ we use a value of $550$, the mean of the values found by \citet{Abel2016} for components A and B of the Veil based on the properties of the Trapezium stars \citep{Ferland2012} and their relative distances, $2$~pc and $4.2$~pc.
This $G_{0}$ value should be valid for most of the gas in the Veil as this structure is a spherical shell \citep[][]{Pabst2019}.
Using this value of $G_{0}$ and the derived gas properties we have that $\gamma=(3\mbox{--}6)\times10^{3}$~K$^{1/2}$~cm$^{3}$.
A comparison between the heating efficiency as a function of $\gamma$ measured towards different regions is presented in Figure~\ref{fig:heat}.
The overall picture is that the theoretical predictions of the heating efficiency overpredict the observed values.
This discrepancy might mean that the heating efficiency is lower, that the PAH abundance is lower, or that there is a bias in the observed values due to the use of TIR as an estimate of the FUV radiation field \citep[e.g.,][]{Hollenbach1999,Okada2013,Kapala2017}.
Here we do not investigate this  further.

\subsection{Background molecular cloud; Orion~A}

Here we use the C$30\alpha$, C$65\alpha$, C$91\alpha$, and $158$~$\mu$m-[CII] lines to study the gas properties in the dense PDR in the envelope of Orion~A.
Emission from these lines at a velocity of $\approx9$~km~s$^{-1}$ is associated with the background molecular cloud.

\subsubsection{CRRLs}
\label{sssec:densepdrcrrls}

The C$30\alpha$ cube overlaps with the observations of C$65\alpha$ and C$91\alpha$ of \citet{Wyrowski1997} (see Figure~\ref{fig:c30mom0}).
Here we use the ratios between the intensities of these lines to constrain the gas properties.
Since the lines trace the PDR at the interface between the HII region and the background molecular cloud, the background continuum will be zero \citep[e.g.,][]{Natta1994}.

\begin{figure}
 \includegraphics[width=0.49\textwidth]{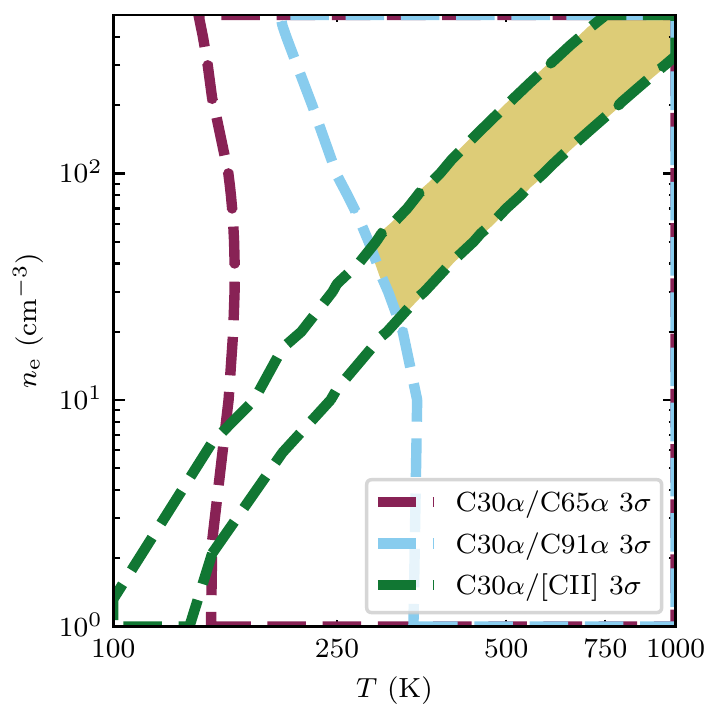}
 \caption{Constraints on the gas temperature and electron density imposed by the ratios between the C$30\alpha$, C$65\alpha$, C$91\alpha$ and $158$~$\mu$m-[CII] lines.
          The yellow shaded region shows where the constraints overlap.
          \label{fig:crrl_ratio_ghz}}
\end{figure}

We focus on a $40\arcsec$ region to the north of Orion-KL, at $(\alpha,\delta)_{\mathrm{J2000}}=(05^{\mathrm{h}}35^{\mathrm{m}}16.7828^{\mathrm{s}},-05\degr22^{\mathrm{m}}02.7225^{\mathrm{s}})$.
There the C$30\alpha$, C$65\alpha$, and C$91\alpha$ cubes overlap, and \citet{Wyrowski1997} provides measurements of the C$65\alpha$ and C$91\alpha$ intensity.
We estimate the error on the intensity of the C$91\alpha$ line from the profile shown in Figure~2 of \citet{Wyrowski1997}.
The root mean square (rms) of the spectrum is close to $0.05$~K, and given that the line profile is narrow and shows little contribution from the HeRRL, we estimate an error of $0.1$~km~s$^{-1}$ on the line width.
These values imply a $1\sigma$ error of $0.2$~K~km~s$^{-1}$ for a $2.9$~K~km~s$^{-1}$ intensity.
For the C$65\alpha$ line we adopt an error of $20\%$ of the observed line intensity.
The C$30\alpha$ line intensity over this region is $71\pm13$~mK~km~s$^{-1}$.

In the studied region, the C$30\alpha$/C$65\alpha$ line ratio is $0.12\pm0.02$ and the C$30\alpha$/C$91\alpha$ line ratio $0.038\pm0.005$.
The constraints imposed on the gas temperature and density by these ratios are shown in Figure~\ref{fig:crrl_ratio_ghz}.
The temperature is constrained to values higher than $150$~K, but they do not constrain the electron density.
The C$65\alpha$/C$91\alpha$ line ratio is $0.30\pm0.06$, and, given the adopted errors, it does not constrain the gas properties.

To fully exploit the power of CRRLs, to provide independent constraints on the gas properties, higher signal-to-noise detections of the observed lines are required.
For example, if the error on the intensity of the C$65\alpha$ line was $10\%$ of the observed value and that of the C$30\alpha$ a factor of two lower, then it would be possible to determine the gas temperature and density using only CRRLs.
Under this assumption, the gas temperature would be constrained to within $10$~K and the electron density within $45$~cm$^{-3}$.
Alternatively, we could use CRRLs at lower frequencies.
At lower frequencies the frequency separation between adjacent C$n\alpha$ lines decreases, hence it becomes easier to achieve higher signal-to-noise ratios by stacking.
Higher resolution observations are also important as they make it  possible to observe the layered structure on higher density PDRs.

\subsubsection{CRRLs and FIR [CII] line}
\label{sssec:densepdrcii}

When the $158$~$\mu$m-[CII] line is optically thick its ratio relative to a CRRL depends on the C$^{+}$ column density, and thus we need an independent measure of the column density to compare them.
To determine the C$^{+}$ column density we use the [$^{13}$CII] $F=2\mbox{--}1$ line.
This line has a velocity difference of $11.2$~km~s$^{-1}$ with respect to the $158$~$\mu$m-[CII] line.
To estimate the column density from $158$~$\mu$m-[CII] and its isotopologue we follow the analysis of \citet{Goicoechea2015}. 
We adopt the corrected line strengths of \citet{Ossenkopf2013} for the three [$^{13}$CII] hyperfine structure lines and a [C/$^{13}$C] abundance ratio of $67$ \citep{Langer1990}, and compute the excitation temperature assuming that the $158$~$\mu$m-[CII] line is optically thick.

For the region studied previously in CRRLs ($(\alpha,\delta)_{\mathrm{J2000}}$ $=(05^{\mathrm{h}}35^{\mathrm{m}}16.7828^{\mathrm{s}},-05^{\mathrm{\circ}}22^{\mathrm{m}}02.7225^{\mathrm{s}})$), we have peak line temperatures of $177$~K and $4$~K for [CII] and [$^{13}$CII] $F=2\mbox{--}1$, respectively.
This translates to an optical depth of $2.3$.
For a background temperature of $35$~K, the excitation temperature of the $158$~$\mu$m-[CII] line is $230$~K.
Using the observed full width at half maximum of $\approx4$~km~s$^{-1}$ this corresponds to a [CII] column density of $9.7\times10^{18}$~cm$^{-2}$.

With an estimate of the [CII] column density we can use the ratio between the $158$~$\mu$m-[CII] line and the CRRLs to further constrain the gas properties.
For the C$30\alpha/$[CII] ratio we have a value of $(1.4\pm0.2)\times10^{-4}$.
The C$30\alpha/$[CII] ratio puts a constraint on the gas properties of the form $n_{\mathrm{e}}\propto T^{3}$, shown in Figure~\ref{fig:crrl_ratio_ghz} with green dashed lines.
Using the lower frequency CRRLs or the [$^{13}$CII] $F=2\mbox{--}1$ line results in a similar constraint.

\subsubsection{Combined constraints}
\label{sssec:pdrgas}

As seen in Figure~\ref{fig:crrl_ratio_ghz} the constraints imposed by the CRRL and $158$~$\mu$m-[CII] line ratios overlap for temperatures higher than $300$~K and an electron density higher than $25$~cm$^{-3}$.
If we assume that all the free electrons come from the ionization of carbon, and a carbon abundance with respect to hydrogen of $1.4\times10^{-4}$, this sets a lower limit to the gas thermal pressure of $5\times10^{7}$~K~cm$^{-3}$.
This is similar to the thermal pressure for the atomic gas layers found by \citet{Goicoechea2016} towards the Orion Bar.

\subsubsection{PDR models}

Motivated by the resemblance between the observed gas distribution (Figure~\ref{fig:slices}) and the structure seen in a PDR (Figure~\ref{fig:pdr}), we compare the observed line intensities to the predictions of PDR models.
In a PDR close to face-on the C$^{+}$ column density is determined by the radiation field and gas density, hence we do not need an independent estimate of the column density.
The PDR models also take into account the gas density and temperature structure.

We focus on the region previously studied in Sect.~\ref{sssec:densepdrcrrls}, towards the north of Orion-KL.
To make a comparison with the PDR model predictions, we need to take into account the geometry;  if the PDRs are not observed face-on, then the column density along the line of sight is not determined by the radiation field and density.
For example, the Orion Bar has a length of $0.28\pm0.06$~pc along the line of sight \citep{Salgado2016}, while in the perpendicular direction its extent is $\approx0.02$~pc \citep[e.g.,][]{Wyrowski1997,Goicoechea2016}.
To determine the length of the PDR along the line of sight we use the intensity of the C$30\alpha$ line, then we use this to scale the rest of the line intensities.
Once we have scaled the line intensities, we determine which models are able to reproduce the observed line intensities and ratios.

First we make a comparison with constant density PDR models.
These models require densities higher than $5\times10^{5}$~cm$^{-3}$ to explain the line intensities and ratios.
This is equivalent to an electron density higher than $70$~cm$^{-3}$, which is consistent with the values found towards this region (Figure~\ref{fig:crrl_ratio_ghz}); however, these models also require radiation fields $G_{0}\geq5\times10^{5}$.
In this region, which is a factor of $2.5$ closer to the Trapezium than the Orion Bar, the incident radiation field should be a factor of six larger than in the Bar, or $G_{0}\approx1.4\times10^{5}$.
This shows that constant density PDR models are not able to explain the observed line properties given reasonable input parameters.

Next we make a comparison with stationary isobaric PDR models.
In this case the models require thermal pressures larger than $5\times10^{7}$~K~cm$^{-3}$, and a radiation field $G_{0}=(0.4\mbox{--}1)\times10^{5}$ to explain the observations.
The isobaric model that best reproduces the observations has $G_{0}=1\times10^{5}$ and $P_{\mathrm{th}}=5\times10^{7}$~K~cm$^{-3}$.
In this case the radiation field and gas thermal pressure are consistent with independent estimates.
Stationary isobaric PDR models also provide better results when explaining observations of excited molecular tracers \citep[e.g.,][]{Joblin2018}.

Given the best fit isobaric PDR model, we assess whether the constraints derived assuming a homogeneous gas slab are reasonable.
In this model the CRRL emission originates mostly from a layer with a gas temperature of $200$~K, and a similar excitation temperature for the $158$~$\mu$m-[CII] line.
The gas temperature is $30\%$ lower than that derived under the homogeneous slab model ($300$~K).
The electron density in the isobaric PDR model is $75$~cm$^{-3}$ in the layer where the CRRL emission peaks, i.e., roughly $50\%$ higher than in the homogeneous slab model.
Therefore, the lower limits from the homogeneous model predict a gas thermal pressure which is $25\%$ to $50\%$ lower than that predicted by a stationary isobaric PDR model.
This difference is not significant considering that the isobaric PDR models used only sparsely sample the $P_{\mathrm{th}}$--$G_{0}$ space.

\section{Summary}
\label{sec:summary}

We  presented CRRL observations towards Orion~A in the frequency range $230$--$0.15$~GHz, including the first detections of the lines in absorption, with the aim of comparing them with the $158$~$\mu$m-[CII] line.
The CRRLs towards Orion~A show the presence of multiple velocity components, similar to what is observed through other tracers of neutral gas (e.g., the $21$~cm-HI line or the $158$~$\mu$m-[CII] line).
We identify CRRL emission associated with the Veil and the background molecular cloud.

We find that the Veil is preferentially traced using lines at frequencies $\lesssim2$~GHz ($n>150$ for C$n\alpha$ lines), similar to the findings of previous studies,  because M42 becomes opaque at these frequencies and its continuum produces significant amplification of the foreground lines.
Using C$280\alpha$/C$351\alpha$ and C$350\alpha$/[CII] line ratios we were able to constrain the properties of the Veil on $36\arcmin$ scales ($4.3$~pc).
We find a gas temperature of $30$~K$\leq T_{\mathrm{e}}\leq45$~K and an electron density of $0.65$~cm$^{-3}\leq n_{\mathrm{e}}\leq0.95$~cm$^{-3}$, where the quoted ranges consider $3\sigma$ errors on the line ratios.
From these physical conditions we constrain the gas cooling rate through the $158$~$\mu$m-[CII] line and the efficiency of photoelectric heating.
We find a lower limit on the gas cooling rate of $(4\pm0.2)\times10^{-26}$~erg~s$^{-1}$~(H-atom)$^{-1}$, and a photoelectric heating efficiency of $\epsilon_{\mathrm{pe}}=(6.9\pm0.3)\times10^{-3}$ for $\gamma=(3\mbox{--}6)\times10^{3}$~K$^{1/2}$~cm$^{3}$.
Based on these values, the Veil is classified somewhere between a diffuse cloud and a dense PDR.

The dense PDR, at the interface between the HII region and Orion~A, is traced using CRRLs at frequencies $\gtrsim2$~GHz.
By comparing the spatial distribution of the C$30\alpha$ and C$65\alpha$ lines to that of the $158$~$\mu$m-[CII] and $^{12}$CO$(2\mbox{--}1)$ lines, we find a clump to the south of the Trapezium where we can observe the layered PDR structure.
The relative location of the C$65\alpha$ line with respect to the $158$~$\mu$m-[CII] line indicates that in a dense PDR the radio lines trace colder gas than the FIR line.

Motivated by the observed distribution of atomic and molecular gas tracers we compared the intensity of the CRRLs and the $158$~$\mu$m-[CII] line to the predictions of PDR models.
We find that stationary isobaric PDR models are able to reproduce the observations.
They imply a thermal gas pressure $\geq5\times10^{7}$~K~cm$^{-3}$, and likely a factor of two higher.
This result agrees with the thermal pressure derived from CRRL and $158$~$\mu$m-[CII] line ratios.

This work shows that the combined use of CRRLs and the $158$~$\mu$m-[CII] line is a powerful tool for   studying the ISM.
They provide an alternative method to determine the gas physical conditions (temperature and electron density).
The physical conditions derived this way can be combined with the information provided by the $158$~$\mu$m-[CII] line to determine the gas heating and cooling.

With new and upgraded telescopes (e.g., SKA, ngVLA, uGMRT, LOFAR2.0), CRRLs will allow us to explore the ISM in our Galaxy and others \citep[][]{Morabito2014,Emig2018}.
In distant galaxies, where spectral tracers of the ISM may be harder to come by, the use of CRRLs and the $158$~$\mu$m-[CII] line can provide important constraints on the properties of the ISM across cosmic time.
Our work shows how these constraints can be obtained by taking into consideration the structure of a PDR.

\begin{acknowledgements}
The authors would like to thank the anonymous referee for the feedback.
 P.S., J.B.R.O., K.L.E., A.G.G.M.T., and H.J.A.R. acknowledge financial support from the Dutch Science Organisation (NWO) through TOP grant 614.001.351.
A.G.G.M.T. acknowledges support through the Spinoza premie of the NWO.
LOFAR, designed and constructed by ASTRON, has facilities in several countries, that are owned by various parties (each with their own funding sources), and that are collectively operated by the International LOFAR Telescope (ILT) foundation under a joint scientific policy.
This paper makes use of the following ALMA data: ADS/JAO.ALMA\#2013.1.00546.S. ALMA is a partnership of ESO (representing its member states), NSF (USA), and NINS (Japan), together with NRC (Canada), MOST and ASIAA (Taiwan), and KASI (Republic of Korea), in cooperation with the Republic of Chile. The Joint ALMA Observatory is operated by ESO, AUI/NRAO, and NAOJ.
Based in part on observations made with the NASA/DLR Stratospheric Observatory for Infrared Astronomy (SOFIA). SOFIA is jointly operated by the Universities Space Research Association, Inc. (USRA), under NASA contract NNA17BF53C, and the Deutsches SOFIA Institut (DSI) under DLR contract 50 OK 0901 to the University of Stuttgart.
Partly based on observations with the 100m telescope of the MPIfR (Max-Planck-Institut für Radioastronomie) at Effelsberg.
Part of this work was carried out on the Dutch national e-infrastructure with the support of the SURF Cooperative through grant e-infra 160022 \& 160152.
This research has made use of the SIMBAD database, operated at CDS, Strasbourg, France, and of the NASA/IPAC Infrared Science Archive, which is operated by the Jet Propulsion Laboratory, California Institute of Technology, under contract with the National Aeronautics and Space Administration.
This research made use of Astropy, a community-developed core Python package for Astronomy \citep{Astropy2013}, and matplotlib, a Python library for publication quality graphics \citep{Hunter2007} and LMFIT, a nonlinear least-squares minimization and curve-fitting package for Python \citep{newville_2014_11813}.
Data from the literature was digitized using WebPlotDigitizer version $4.1$ when not available in digital form \citep{webplot}.
Analysis of CRRL observations was done using CRRLpy \citep{crrlpy}.
\end{acknowledgements}

\bibliography{ref_rrl.bib}

\appendix

\section{Nonlinear gain correction}
\label{app:nlg}

Under ideal circumstances, the relation between the raw counts measured by a radio telescope $P$, the source temperature $T_{\mathrm{sou}}$, and the system temperature $T_{\mathrm{sys}}$ will be of the form
\begin{equation}
 P^{\mathrm{[CAL]}}=G(T_{\mathrm{sou}}+T_{\mathrm{sys}}^{\mathrm{[CAL]}})+C,
 \label{eq:lg}
\end{equation}
where $G$ is the conversion factor between temperature and telescope units (counts) and $C$ a constant offset between the two scales.
Here we  adopted the nomenclature of \citet{Winkel2012}, in which $T_{\mathrm{sys}}^{\mathrm{[CAL]}}$ denotes the system temperature, considering the possible contribution from a calibration signal $T_{\mathrm{cal}}$, and $T_{\mathrm{sou}}$ is the temperature of the astronomical source of interest which includes both continuum and line, $T_{\mathrm{cont}}$ and $T_{\ell}$, respectively.
In order to determine the conversion between counts and temperature, procedures such as those outlined in \citet{Winkel2012} were used.

Given the nature of the signal path on a radio telescope, it is possible that relation~\ref{eq:lg} will break down.
This could be due to the amplifiers being driven out of their linear response regime (e.g., if a bright source is observed).
This has the effect of changing Equation~\ref{eq:lg} to
\begin{equation}
 P^{\mathrm{[CAL]}}=G(T_{\mathrm{sou}}+T_{\mathrm{sys}}^{\mathrm{[CAL]}})+G_{\mathrm{nl}}(T_{\mathrm{sou}}+T_{\mathrm{sys}}^{\mathrm{[CAL]}})^{2}+C.
 \label{eq:nlg}
\end{equation}
Here, $G_{\mathrm{nl}}$ represents the nonlinear contribution of the amplifier gain.

If Equation~\ref{eq:lg} is no longer valid, and we can represent the conversion between raw counts and temperature using Equation~\ref{eq:nlg}, then it is possible to calibrate the raw counts if we make some assumptions about $T_{\mathrm{sou}}$.
To estimate $G_{\mathrm{nl}}$ we can use a reference position $P_{\mathrm{ref}}$, ideally devoid of any astronomical signal, and a model of $T_{\mathrm{sou}}$.
Then,
\begin{equation}
 G_{\mathrm{nl}}=\frac{P_{\mathrm{sou}}-P_{\mathrm{ref}}-GT_{\mathrm{sou}}}{2T_{\mathrm{sys}}+T_{\mathrm{sou}}^{2}}
\end{equation}

If we are interested in recovering the temperature of a spectral line,  we can work with the continuum subtracted spectra $P_{\ell}^{\mathrm{[CAL]}}=P^{\mathrm{[CAL]}}-P_{\mathrm{cont}}^{\mathrm{[CAL]}}$.
The continuum $P_{\mathrm{cont}}^{\mathrm{[CAL]}}$ can be estimated from line-free channels.
The line brightness temperature $T_{\ell}$ is then obtained from
\begin{equation}
 T_{\ell}=P_{\ell}^{\mathrm{[CAL]}}\left[G+G_{\mathrm{nl}}(T_{\ell}+2T_{\mathrm{cont}}+2T_{\mathrm{sys}}^{\mathrm{[CAL]}})\right]^{-1}
\end{equation}

\section{Gaussian fits to RRL spectra}

A decomposition of the spectra presented in Figure~\ref{fig:gbt2lbandcrrls} into Gaussian components is tabulated in Table~\ref{tab:gbtcrrlsgfit}.

\begin{table}[!h]
\begin{center}
\caption{Best fit Gaussian parameters for RRLs observed towards M42}
\begin{tabular}{lccc}
\toprule
Region & $v_{\mathrm{lsr}}$ & $T_{\mathrm{mb}}$ & $\Delta v$ \\
       & (km s$^{-1}$)  & (K)           & (km s$^{-1}$) \\
\midrule
 H$137\alpha$ & $-6.92\pm0.05$ &  $3.95\pm0.01$ & $32.9\pm0.1$ \\
 H$145\alpha$ & $-7.57\pm0.08$ &  $2.98\pm0.01$ & $34.7\pm0.2$ \\
 H$151\alpha$ & $-8.5\pm0.1$ &  $4.40\pm0.03$ & $34.8\pm0.3$ \\
 H$155\alpha$ & $-8.2\pm0.1$ &  $2.21\pm0.01$ & $36.0\pm0.2$ \\
 H$156\alpha$ & $-7.8\pm0.1$ &  $2.85\pm0.02$ & $36.1\pm0.3$ \\
 H$164\alpha$ & $-7.9\pm0.1$ &  $2.26\pm0.02$ & $36.0\pm0.4$ \\
 H$174\alpha$ & $-8.3\pm0.2$ &  $1.56\pm0.02$ & $37.0\pm0.5$ \\
 H$280\alpha$ &  $0\pm2$ &  $0.010\pm0.002$ & $25\pm5$ \\

He$137\alpha$ & $-5\pm1$  &  $0.42\pm0.02$ & $31\pm2$ \\
He$145\alpha$ & $-11\pm2$ &  $0.25\pm0.04$ & $23\pm3$ \\
He$151\alpha$ & $-6\pm1$  &  $0.40\pm0.05$ & $15\pm4$ \\
He$155\alpha$ & $-10\pm3$ &  $0.17\pm0.02$ & $24\pm5$ \\
He$156\alpha$ & $-7\pm5$  &  $0.15\pm0.04$ & $16\pm10$ \\
He$164\alpha$ & $-7\pm3$  &  $0.20\pm0.03$ & $17\pm6$ \\
He$174\alpha$ & $-14\pm2$ &  $0.19\pm0.03$ & $32\pm4$ \\

 C$137\alpha$ &  $7.3\pm0.6$ &  $0.26\pm0.05$   & $12\pm2$ \\
 C$145\alpha$ &  $6.3\pm0.7$ &  $0.24\pm0.06$   &  $9\pm2$ \\
 C$151\alpha$ &  $5.2\pm0.8$ &  $0.67\pm0.07$   & $10\pm1$ \\
 C$155\alpha$ &  $4.4\pm0.4$ &  $0.31\pm0.06$   &  $9\pm1$ \\
 C$156\alpha$ &  $6\pm1$     &  $0.5\pm0.1$     & $10\pm1$ \\
 C$164\alpha$ &  $4.9\pm0.7$ &  $0.54\pm0.07$   & $10\pm1$ \\
 C$174\alpha$ &  $5.1\pm0.3$ &  $0.47\pm0.05$   &  $6\pm1$ \\
 C$280\alpha$ &  $0.7\pm1.0$ & $-0.023\pm0.003$ & $11\pm1$ \\
\bottomrule
\end{tabular}
\label{tab:gbtcrrlsgfit}
\end{center}
\end{table}

\end{document}